\documentclass[preprint]{aastex631}
\usepackage{multirow}
\usepackage{color}
\usepackage{graphicx}
\usepackage{subfigure}
\usepackage{natbib}
\usepackage{booktabs}
\usepackage{chngcntr}

\usepackage[encapsulated]{CJK}
\usepackage{mathrsfs}
\usepackage{amsmath}
\usepackage{epstopdf}
\usepackage{longtable}
\usepackage{booktabs}
\usepackage{hyperref} 

\newbox\grsign \setbox\grsign=\hbox{$>$} \newdimen\grdimen \grdimen=\ht\grsign
\newbox\laxbox \newbox\gaxbox
\setbox\gaxbox=\hbox{\raise.5ex\hbox{$>$}\llap
     {\lower.5ex\hbox{$\sim$}}}\ht1=\grdimen\dp1=0pt
\setbox\laxbox=\hbox{\raise.5ex\hbox{$<$}\llap
     {\lower.5ex\hbox{$\sim$}}}\ht2=\grdimen\dp2=0pt

\shorttitle{census of CO completeness}
\shortauthors{Sun et al.}

\newcommand{\s}{\,{\rm s}}

        \newcommand{\K}{\,{\rm K}}
\newcommand{\co}{$^{12}$CO}                             
\newcommand{\xco}{$^{13}$CO}                            
\newcommand{\xxco}{C$^{18}$O}                           
\newcommand{\kms}{km\,s$^{\rm -1}$}

\newcommand{\kkmsarc}{K\,km\,s$^{\rm -1}$\,arcmin$^{\rm 2}$}
\newcommand{\vlsr}{$V_{\rm LSR}$}
\graphicspath{{./}{figures/}}
\begin{document}

\title{Examinations of CO completeness based on three independent CO surveys}
\correspondingauthor{Ji Yang \& Yan Sun}
\email{jiyang@pmo.ac.cn, yansun@pmo.ac.cn}

\author[0000-0002-3904-1622]{Yan Sun}\affiliation{Purple Mountain Observatory, Chinese Academy of Sciences, Nanjing 210008, China}

\author[0000-0001-7768-7320]{Ji Yang}
\affiliation{Purple Mountain Observatory, Chinese Academy of Sciences, Nanjing 210008, China}

\author[0000-0003-4586-7751]{Qing-Zeng Yan}
\affiliation{Purple Mountain Observatory, Chinese Academy of Sciences, Nanjing 210008, China}

\author{Zehao Lin}
\affiliation{Purple Mountain Observatory, Chinese Academy of Sciences, Nanjing 210008, China}
\affiliation{University of Science and Technology of China, Hefei, Anhui 230026, China}

\author[0000-0003-2549-7247]{Shaobo Zhang}
\affiliation{Purple Mountain Observatory, Chinese Academy of Sciences, Nanjing 210008, China}

\author[0000-0002-0197-470X]{Yang Su}
\affiliation{Purple Mountain Observatory, Chinese Academy of Sciences, Nanjing 210008, China}

\author{Ye Xu}
\affiliation{Purple Mountain Observatory, Chinese Academy of Sciences, Nanjing 210008, China}

\author{Xuepeng Chen}
\affiliation{Purple Mountain Observatory, Chinese Academy of Sciences, Nanjing 210008, China}

\author[0000-0003-0746-7968]{Hongchi Wang}
\affiliation{Purple Mountain Observatory, Chinese Academy of Sciences, Nanjing 210008, China}

\author[0000-0003-2418-3350]{Xin Zhou}
\affiliation{Purple Mountain Observatory, Chinese Academy of Sciences, Nanjing 210008, China}


\accepted{to \apjs\ July 1, 2021}

\begin{abstract}
We report the global properties recovered by an ongoing CO survey of the Milky
Way Imaging Scroll Painting (MWISP) toward the Galactic outskirts. 
Our results are also compared to those extracted by a uniform decomposition method from 
the CfA 1.2 m CO survey and the FCRAO 14 m outer Galaxy survey (OGS).
We find that more extended and unseen structures are present in the MWISP data.
The total flux across the disk recovered by the MWISP survey is 1.6 times larger than those
recovered by the CfA and OGS surveys in the case of the same resolution. The discrepancies are scaling with distance. For example, 
in the outermost OSC arm, the flux ratios for MWISP-to-CfA and MWISP-to-OGS increase up to 43.8 and 7.4, respectively.   
Nonetheless, the census of molecular gas in our Galaxy is still far from complete by the MWISP, with flux
completeness of $<$58\%. The total mass ratios of the tabulated molecular clouds between  
different surveys are similar to the CO flux ratio. 
The application of these ratios to the total H$_{\rm 2}$ mass of 
our Galaxy yields a correction factor of at least 1.4, meaning that the H$_{\rm 2}$ mass of our Galaxy 
should be at least 40\% more massive than previously determined.
Including the completeness correction, an even more significant fraction
of the matter should be contributed by baryonic matter. 
The mass spectrum in the outer Galactic plane is better described by a non-truncating power-law with $\gamma$=$-$1.83$\pm$0.05, 
and an upper mass of $M_0$=(1.3$\pm$0.5)$\times$10$^{\rm 6}$~$M_{\sun}$.  
\end{abstract}


\section{Introduction} \label{sec:intro}

The $J$ = 1$\rightarrow$0 line of \co~ is the most accessible and widely used tracer of the molecular interstellar medium 
due to the combination of high abundance and low permanent dipole moment~\citep{bolatto2013,heyer2015}.
As a consequence, most scientific time of the single-dish telescopes~(e.g., the 1.2 m, the NANTEN 4 m, the 
Bell Laboratories 7 m, and the FCRAO 14 m telescopes) was devoted to the CO survey projects
~\citep[e.g., ][]{heyer1998,dame2001,lee2001,nanten2004,grs2006} 
since the discovery of the CO line in 1970. 
The major scientific achievements of the large CO surveys were summarized in a recent review by \citet{heyer2015}. 
These CO surveys have been proved essential for improving our knowledge of the physics of 
molecular ISM, the physics of star formation, as well as the structure and dynamics of the Milky Way.

Benefit from the increasing of sensitivity, bandwidth, spectral channels, and enlarging of field of view by 
applying multi-beam receivers, more moderate- to large-size 
single-dish telescopes became capable of wide-field imaging of CO and its isotopologues, \xco~ and \xxco. 
There are currently several new surveys being undertaken with 
new instruments equipped on the Mopra 22 m~\citep[e.g., ][]{barnes2015,barnes2018}, 
the Nobeyama 45 m~\citep[e.g., ][]{fugin2017}, and the Purple Mountain Observatory (PMO) 
13.7 m telescopes~\citep[e.g., ][]{su2019}.
These new surveys allow at the same time high resolution, high sensitivity, and wide velocity coverage, 
enabling the examination of the internal structure of the clouds  
and the discovery of the faint and extended cloud envelopes, as well as the discovery of 
many new molecular clouds~(MCs) with small angular size, and/or low surface brightness. 
Indeed, using the data from the ongoing project Milky Way Imaging Scroll Painting (MWISP)
with the PMO 13.7 m telescope, we have detected many new MCs, and a new segment of the distant spiral 
arm~\citep[e.g.,][]{sun2015,su2016,du2016,sun2017}, and have revealed the complex distribution of molecular material 
within the parent cloud~\citep[e.g.,][]{zhang2014,su2020,lin2020}. 

While much scientific progress has been achieved from the analysis of CO survey data, 
our knowledge of the completeness of these CO surveys is still very limited.
The observational effects on the cloud properties have long been known.
For example, \citet{rosolowsky2006} 
reported that the bias of measurements due to the observational effects is at least 40\%.
Based on the MWISP data ($l$ = [25\fdg8, 49\fdg7], and $b$=[$-$5\fdg25, +5\fdg25]), 
\citet{yan2021b} investigated the dependence of observed properties of MCs (i.e., brightness temperature, flux) 
on angular resolution (beam sizes) and sensitivity by simulating observations with larger beam sizes and lower sensitivities.
They found that the observed brightness temperature is mainly controlled by the 
resolution~(beam size) of the observation. 
Rather low filling factors are reported for small clouds, i.e., only $\sim$50\% for cloud with
size of two resolution elements~\citep{yan2021b}. Furthermore, the flux completeness of a cloud strongly depends 
on the mean signal-to-noise ratio (S/N) of the cloud, which is about 0.5 for the case of S/N=3.3~\citep{yan2021b}. 
All these findings strongly indicate that the census of molecular gas from the CO surveys with finite 
sensitivities and resolutions is mostly incomplete.

Currently, the region within Galactic coordinates $l$=[104\fdg75,150\fdg25] and $b$=[$-$5\fdg25,+5\fdg25]
has been fully covered by the MWISP survey~\citep{yan2021a}, as well as partly covered by the FCRAO outer Galaxy CO
survey \citep[OGS;][]{heyer1998} and the CfA 1.2 m CO survey~\citep[][hereafter the CfA survey]{dame2001}.
Such three independent surveys overlapped over a large area of the sky provide us an opportunity 
to investigate the completeness and limits of the CO surveys.
The second quadrant contains four well-separated spiral arms along the line of sight. Thus it can 
additionally offer advantages over other Galactic quadrants when investigating the effects of distance. 
In this study, we investigate the recovered flux and its completeness measured by the CO surveys. 
This paper is organized as follows. First, the data and signal identification and decomposition methods are presented in Section 2, 
and in Section 3 the integrated cloud properties measured by the MWISP survey are presented along with a comparison with those measured by other surveys. 
 Next, in Section 4 we discuss the lessons from the current CO surveys and the correction factor of total H$_{\rm 2}$ mass of our Galaxy. 
In Section 5 we summarize the results.

\section{Data and analysis} \label{sec:data}
This study is on the basis of the overlapped region of the MWISP, CfA, and OGS surveys, in the region of 
$l$ = [104\fdg75, 141\fdg54], $b$ = [$-$3\fdg028, 5\fdg007], and \vlsr~= [$-$137, 40]~\kms. 

\subsection{MWISP data}
The MWISP project is an unbiased Northern Galactic Plane CO survey, which is capable of simultaneously observing 
J=1-0 \co, \xco, and \xxco~lines~\citep[refer to][etc., for project details and initial results]{zhang2014,sun2015,su2019}. 
The MWISP CO data used here were observed during 2011 and 2020 with the PMO 13.7 m telescope. 
The front end detector was the Superconducting Spectroscopic Array Receiver~(SSAR), a nine-beam, sideband-separating receiver~\citep{shan2012}.
The angular resolution of the 13.7 m telescope is $\sim$50$\arcsec$ at the frequency of the \co(1-0) line.
The on-the-fly (OTF) mode was applied in the MWISP survey. 
The sample spacing between adjacent scans is 15$\arcsec$~\citep[see the details in][for observation setups]{sun}.
Therefore, the MWISP survey fully sampled the mapped area.
The OTF raw data were regridded into 30$\arcsec\times$30$\arcsec$ pixels.
The total velocity coverage is $\sim$2600~\kms~ with a channel width of 0.16~\kms. 
A first-order baseline correction was applied to the MWISP data.
Please refer to \citet{su2019} for details of the observation strategy and data reduction of the MWISP survey.

\begin{figure*}
\centering
\includegraphics[angle=0,scale=0.5,bb=20 20 430 440]{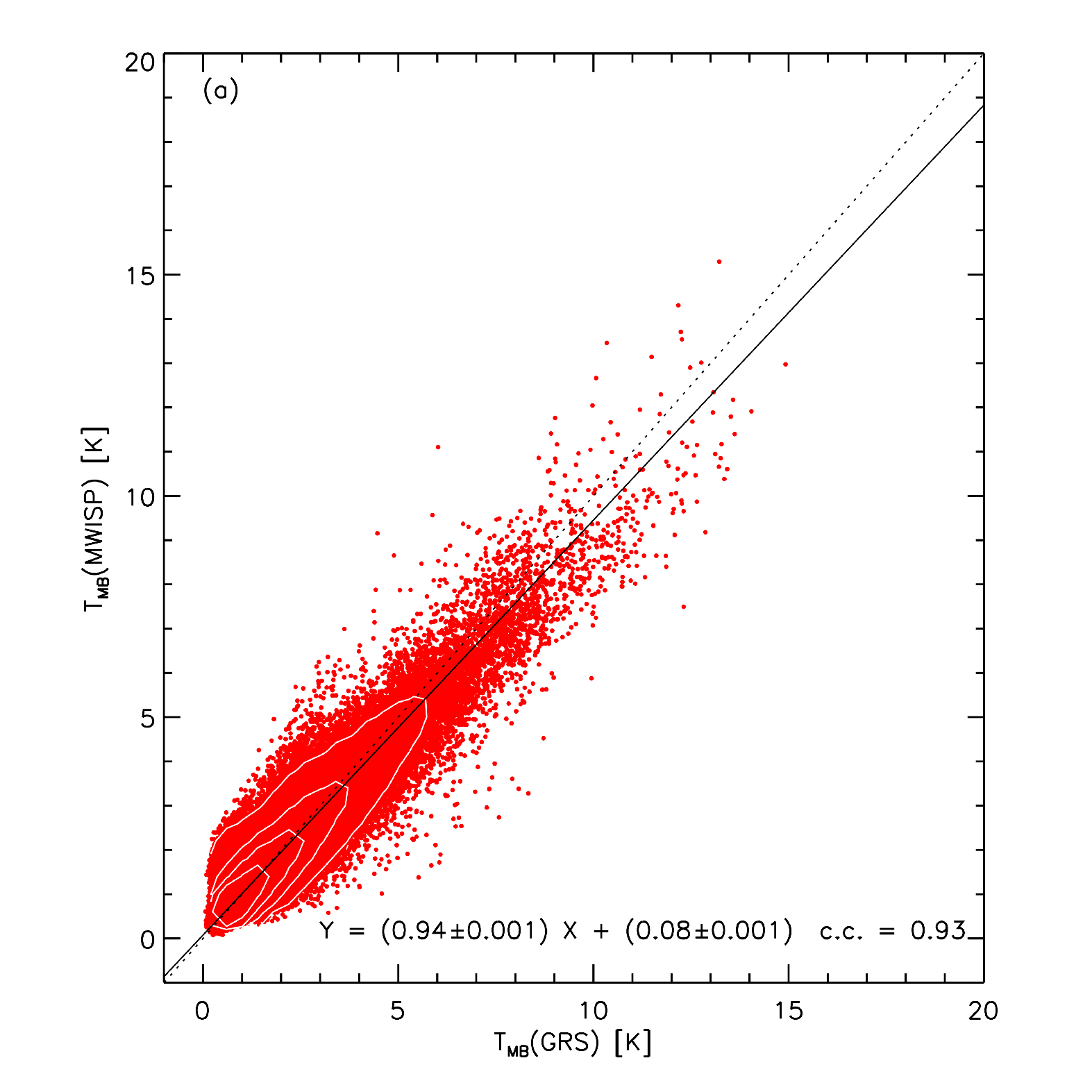}
\includegraphics[angle=0,scale=0.5,bb=20 20 430 440]{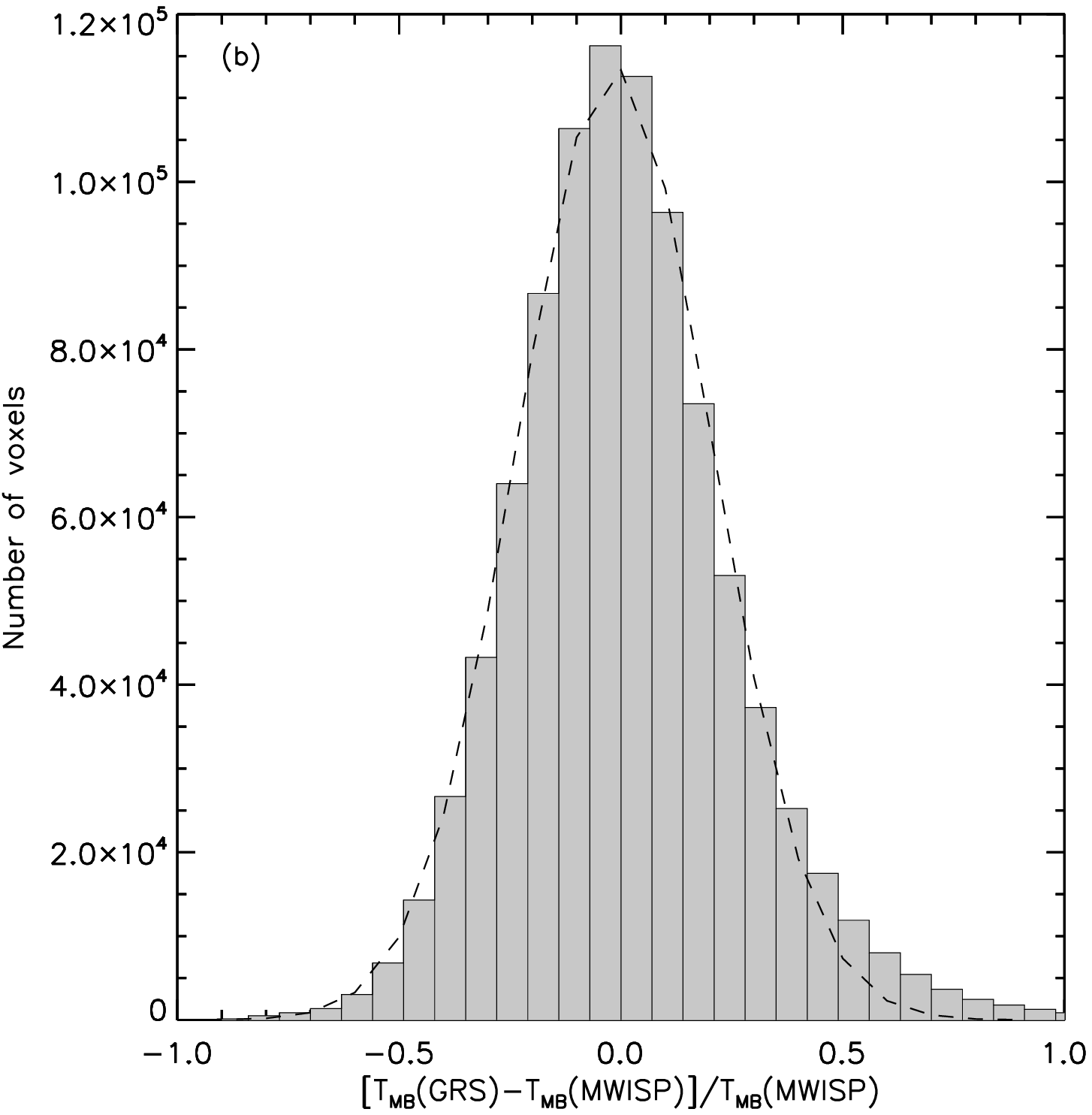}
\caption{The \xco(1-0) intensities from the MWISP survey by the PMO 13.7 m telescope and the GRS by the FCRAO 14 m telescope. 
The compared area is $l$=[29\fdg75, 40\fdg25], $b$=[$-$1\fdg0, +1\fdg0], and \vlsr=[$-$5, 135]~\kms. The two data sets are 
both smoothed to the same channel width of 2~\kms. We apply DBSCAN algotithm to identify signal. All voxels above 6$\sigma$ 
noise level are plotted. ($a$) The contours indicate number density of 2$\times$10$^{\rm 2}$, 2$\times$10$^{\rm 3}$, 
1$\times$10$^{\rm 4}$, and 4$\times$10$^{\rm 4}$, respectively. The solid-line shows the best-fit linear function of 
 $T_{\rm MB}$(MWISP)=(0.94$\pm$0.001)$\times$$T_{\rm MB}$(GRS)+(0.08$\pm$0.001). The dotted-line shows the $T_{\rm MB}$(MWISP)=$T_{\rm MB}$(GRS). 
 ($b$) Histogram of the difference between the valid voxel values of the two surveys. 
 The dashed line shows the best fit of the distribution by a Gaussian function, with an offset of 
 $-$0.014 and a standard deviation of 0.219. 
\label{fig:cal}}
\end{figure*}

The chopper-wheel calibration was adopted to get the antenna temperature, $T_{\rm A}^*$, which has been corrected for 
atmospheric absorption~\citep{kutner1981}. Data presented in this paper are further converted to a scale of main beam brightness 
temperature, using $T_{\rm MB}$ = $T_{\rm A}^*\,$/$\eta_{\rm MB}$, where the main beam efficiency $\eta_{\rm MB}$ is typically 
 0.46 at 115 GHz~(see the status report of the 13.7 m telescope~\footnote{http://www.radioast.nsdc.cn/mwisp.php}).
Here, we verify the calibration accuracy of the MWISP data by comparing it with the FCRAO Galactic ring 
survey~\citep[GRS; ][]{grs2006}. Because these two surveys are both fully sampled and have similar resolution 
and sensitivity, the difference of calibrated main beam temperature ($T_{\rm MB}$) would faithfully reflect the calibration uncertainty. 
Figure~\ref{fig:cal} shows the comparison results for the overlapped region of the 
two surveys, e.g., in the region of $l$=[29\fdg75, 40\fdg25],
$b$=[$-$1\fdg0, +1\fdg0], and \vlsr=[$-$5, 135]~\kms. The two data sets are both smoothed to a same channel width of 
2~\kms. The main beam temperatures of all voxels above 6$\sigma$ are pixel-by-pixel compared, which 
can be well fitted by a linear relation of $T_{\rm MB}$(MWISP)=(0.94$\pm$0.001)$\times$$T_{\rm MB}$(GRS)+(0.08$\pm$0.001), with a 
Pearson correlation coefficient of 0.93 (see Fig.~\ref{fig:cal}a). The distribution of the fractional difference 
between all valid voxels, [$T_{\rm MB}$ (GRS)$-$$T_{\rm MB}$ (MWISP)]/$T_{\rm MB}$ (MWISP) is also plotted as a histogram. 
This distribution is fitted by a Gaussian function, with a peak of $-$0.014 and a width of 0.219. 
These indicate that calibration of the MWISP data is robust, and is consistent with that of the GRS data within $\sim$20\%. 

\subsection{Archival Data}
To date, the CO survey conducted with the CfA 1.2 m telescope is the only one that covers the 
whole Galactic plane~\citep{dame2001}. The CfA CO data used here were 
obtained during 1980 and 2001~\citep{dame2001}. The angular resolution of the 1.2 m telescope is $\sim$8.5$\arcmin$.
We used the raw data set where no interpolation was done\footnote{Available at https://www.cfa.harvard.edu/rtdc/CO.}.
The publicly distributed CfA data are in units of $T_{\rm MB}$, where $\eta_{\rm MB}$ of 0.82 was used for correction.
The pixel size of the whole data set is 7.5$\arcmin$, slightly less than one beam size. In the second Galactic quadrant, 
the velocity coverage of the survey is $\sim$$-$137 to $\sim$50~\kms~ with a channel width of 1.3~\kms.

The OGS CO data used in this study are from \citet{heyer1998}, which were observed during 1994 and 1997
with the FCRAO 14 m telescope.
These data, covering a Galactic coordinates range of $l$ = [102\fdg49, 141\fdg54] and $b$= [$-$3\fdg03, 5\fdg41],  
were sampled every 50$\arcsec$ with an angular resolution of $\sim$46$\arcsec$.
The publicly distributed OGS data are in units of $T_{\rm R}^{*}$, corrected for forward 
spillover and scattering efficiency of the telescope, using $T_{\rm R}^{*}$ = $T_{\rm A}^{*}\,$/$\eta_{\rm FSS}$,
where $\eta_{\rm FSS}$ is 0.7. We transform the OGS data to $T_{\rm MB}$ units, using 
$T_{\rm MB}$ = $T_{\rm R}^*\times$($\eta_{\rm FSS}\,$/$\eta_{\rm MB}$), where $\eta_{\rm MB}$ is 0.45.
The velocity coverage is from $-$153 to 40~\kms~ with a channel width of 0.81~\kms.
 Note that unlike the OTF mapping mode of the MWISP survey, the beam-by-beam mapping 
mode was applied in the CfA and OGS surveys. Therefore, these two surveys did not fully sample
the mapped area.  

\begin{deluxetable*}{ccccccc}
\tabletypesize{\scriptsize}
\setlength{\tabcolsep}{0.04in}
\tablewidth{0pt}
\tablecaption{Summary of the survey data.\label{tab:survey}}
\tablehead{Survey & Beam size & Sampling & \vlsr range & $\delta_{v}$  & rms noise$^{c}$ & $\eta_{MB}$ \\
                  &           &          &  (\kms)  &   (\kms)  & (K) }
\startdata
CfA 1.2 m$^{a}$       & 8.5$\arcmin$  & 7.5$\arcmin$ & [$-$137, +40]        & 1.3    & 0.07 &0.82   \\
OGS$^{b}$             & 46$\arcsec$   & 50$\arcsec$  & [$-$153, +40]        & 0.81   & 0.95 & 0.45   \\
MWISP raw             & 50$\arcsec$   & 30$\arcsec$  & $\sim$[-1400, +1200] & 0.16   & 0.46 & 0.46    \\
... smoothed1     & ...           & 50$\arcsec$  & ...                    & 0.81   & 0.23   & ... \\
... smoothed2     & 8.5$\arcmin$  & 7.5$\arcmin$ & ...                    & 1.3    & 0.04   & ... \\
\enddata
\tablecomments{($a$) \citet{dame2001}. ($b$) \citet{heyer1998}. ($c$) The rms noise are in units of $T_{\rm MB}$}
\end{deluxetable*}

\subsection{Signal Identification and Decomposition} \label{sec:identify}
\begin{figure*}
\vspace{-1mm}
\centering
\includegraphics[angle=0,scale=0.6,bb=20 10 400 340]{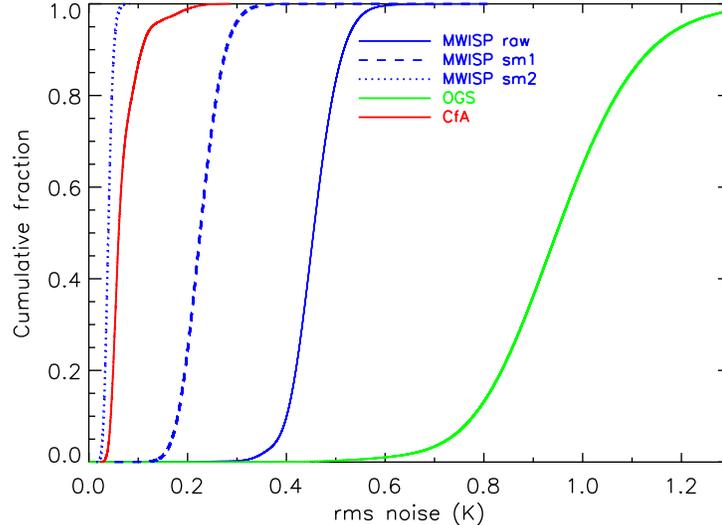}
\caption{The cumulative distributions of $\sigma$ for \co~ emission from the MWISP (blue), OGS (green), and CfA 1.2 m (red) surveys within the
region of $l$=[104\fdg75, 141\fdg54] and $b$=[$-$3\fdg028, +5\fdg007]. The solid lines are derived from the raw data. The dashed-blue
and dotted-blue lines are derived from the smoothed MWISP data with the same resolution and sampling as the OGS
and CfA surveys, respectively.
\label{fig:rms}}
\end{figure*}

We estimate the noises of each data set by measuring the rms intensity, $\sigma_{\rm rms}$,
from the signal-free channels of the data cube. 
The mean values of $\sigma_{\rm rms}$ are 0.46~K, 0.95~K, and 0.07~K for the data sets of the MWISP, OGS, and CfA 
surveys with full resolution, respectively.
The MWISP data are also convolved and 
regridded to match the resolutions and sample spacings of the CfA and OGS surveys. 
The spatial smoothing operation is performed with the \texttt{Python} package 
\texttt{spectral-cube}~\footnote{https://spectral-cube.readthedocs.io/en/latest/index.html}.
The corresponding $\sigma_{\rm rms}$ for the smoothed MWISP data have mean values of 0.04~K, and 0.23~K, 
which are larger than predicted by theory. Further investigations are needed to understand these discrepancies.
The cumulative distributions of $\sigma_{\rm rms}$ for the five data sets of the three CO surveys are shown in Figure~\ref{fig:rms}.

The simplest method for signal identification is ``clipping'', in which all voxels with intensities above some 
statistical significance level are judge to be real signals. But we often find ourselves in
a dilemma when choosing an optimum intensity level. A loose criterion could not significantly suppress the noise, 
while a strict criterion might filter a substantial amount of real emission. Therefore, we adopted a more 
sophisticated approach developed by \citet{yan2020} to identify regions of contiguous, significant emission in 
the position-position-velocity~(PPV) data cubes.  
This approach is based on the density-based spatial clustering of applications with noise (DBSCAN) 
algorithm~\footnote{https://scikit-learn.org/stable/auto\_examples/cluster/plot\_dbscan.html}. 
This algorithm contains three free parameters: (1) MinPts which defines the minimum number of voxels of the structure, 
(2) $\epsilon$ which defines the neighborhood of voxels, and (3) $T_{\rm cutoff}$ which defines the boundary isosurface 
of the structure~\citep{yan2020}. Many combinations of parameters were made before finalizing on MinPts=4,    
$\epsilon$=1, and $T_{\rm cutoff}$=2$\sigma$. Such choice of parameters can identify most of the significant emission (including both the compact
and the extended emission). The definition of connected regions of emission in PPV space here is generally consistent with the trunks 
defined by dendrogram algorithm~\citep{rosolowsky2008}. 
 
 Post selection was also recommended by \citet{yan2020} to ensure including as few noise clouds as 
possible. In this step, a cloud was further rejected from the raw catalogs if 
any one of the following criteria is met: (1) the voxel number below 16, (2) peak 
brightness temperature below 5$\sigma$, (3) projection area has no compact 2$\times$2 region, 
and (4) channel number less than 3. 
We then construct a mask according to the signal identification results, which was applied to the raw data cubes.
The emission-free voxels are blanked. The masked versions of data cubes are used for the following analysis. 
    
\section{Results}\label{sec:results}
\subsection{Large-scale Structure Traced by CO\label{sec:lv}}
\begin{figure*}
\includegraphics[angle=0,scale=0.95,bb=10 80 600 650]{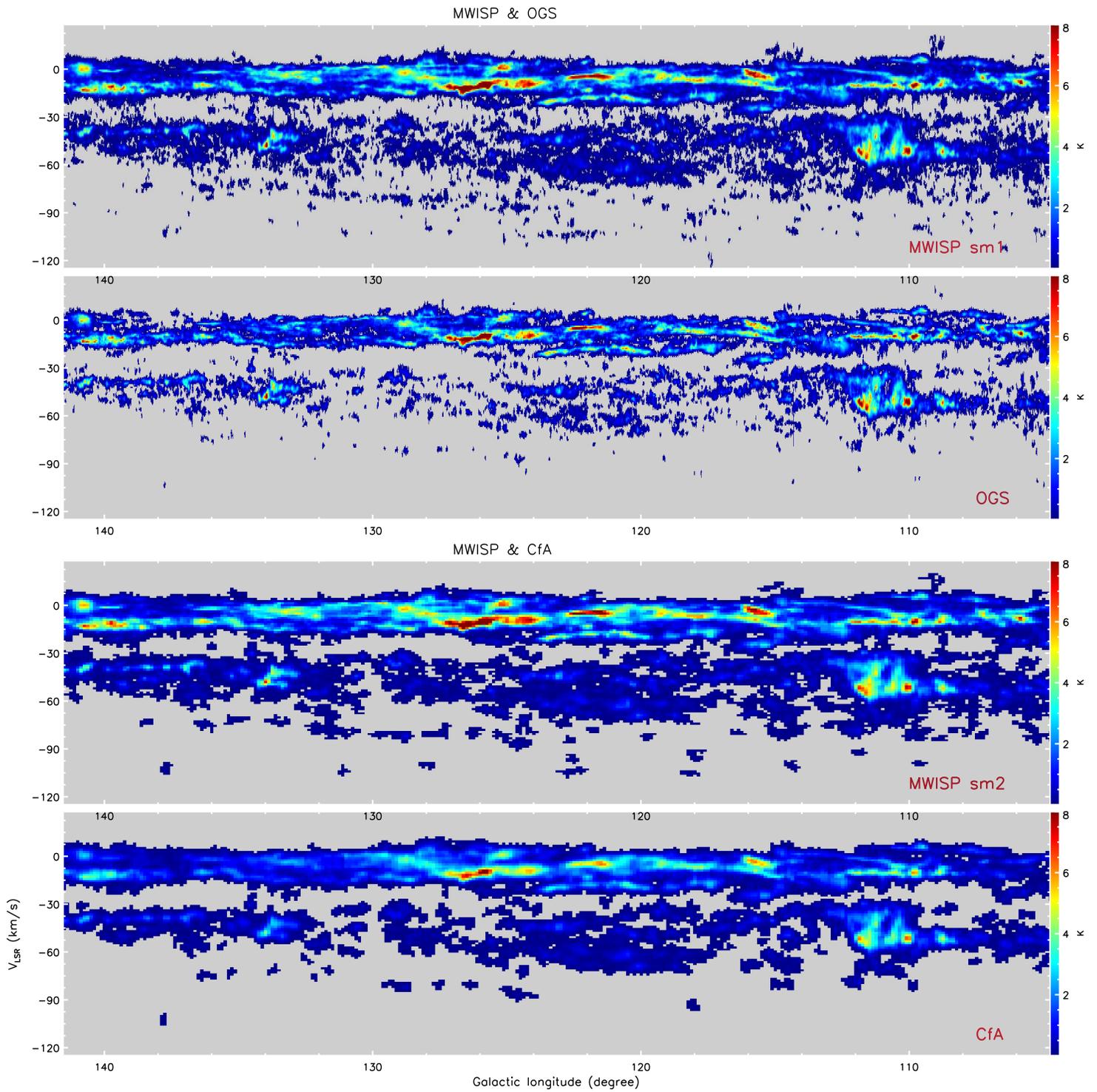}
\caption{The longitude-velocity maps integrated over Galactic latitude range of $b$ = $-$3\fdg028 to 5\fdg007. 
The MWISP data are smoothed to match the resolution and sampling of the OGS and CfA surveys.
The images only contain emission identified by DBSCAN. All emission-free pixels are blanked.
\label{fig:lv}}
\end{figure*}

Figure~\ref{fig:lv} presents the longitude-velocity ($l$-$v$) maps integrated over $b$ = [$-$3\fdg028, 5\fdg007],
which are derived from the noise-suppressed data cubes of the three surveys.
The integrated intensity maps of the four spiral arms derived from the three surveys are also compared in Figure~\ref{fig:intensity}.
Note that the maps derived from the MWISP data with the full resolution and the same resolution as 
the OGS are almost identical by human eyes, thus the maps derived from the former are not displayed here.
Compared with the moment masked $l$-$v$ map of the CfA \citep[Fig. 3 of ][]{dame2001} and the 
unmasked map of the OGS \citep[Fig. 4 of ][]{heyer1998}, the structures shown in their $l$-$v$ maps 
are generally consistent with those revealed in Fig.~\ref{fig:lv}. 
Examined more closely, however, there are some weaker features at the large negative velocity newly recovered 
in Fig.~\ref{fig:lv}. We find that these distant features were actually judged to be real MCs by \citet{digel1994} and \citet{mivi2017}.
The comparison confirms that the DBSCAN performs as well as the often used moment mask method. Moreover, 
our choice of parameters assures a reliable extraction of CO emission, even at a low S/N ratio. 

The features revealed by the three independent surveys are also compared. We find that all data sets with a variety of 
resolutions and sensitivities have the ability to resolve major features of the Local 
arm from $\sim-$25 to $\sim$15~\kms and the Perseus arm from $\sim-$65 to $\sim-$25~\kms, due to their relatively small 
distances and/or large angular sizes. And as expected, the 
internal sub-structure within each arm can only be well resolved by the ones with higher resolution~(e.g., the OGS and MWISP surveys). 
The Outer arm ($\sim-$95 to $\sim-$65~\kms) appears as a more coherent feature in 
the MWISP survey than in the other two surveys. 
Besides, the MWISP data discover a new component lying beyond the Outer arm at the maximum negative 
velocity (at \vlsr$<\sim-$95~\kms, namely the Outer Scutum-Centaurus arm, hereafter OSC), which was 
firstly reported in our previous work~\citep{sun2015}.  
 Some new features with small angular size and/or relatively faint emission are also emerging  
between each pair of adjacent arms by the MWISP survey. 
Considering that these features are extracted by a uniform approach, 
these discrepancies would faithfully reflect the observational effects introduced by the 
finite sensitivity and resolution of each survey. Our lesson from such a rough comparison is that the 
survey with both high sensitivity and resolution will offer complementary or new insights 
into the spiral structure in outer or extreme outer Galaxy regions, 
the internal structure within each spiral arm, and the inter-arm structure between each pair of adjacent arms.

\subsection{Voxel Number and Flux Recovered}
The integrated flux of CO emission represents an essential parameter because it can be well translated into 
total H$_{\rm 2}$ gas mass. Therefore it is investigated in detail in the following two sections. 
The total voxel number can reflect the volume of the CO emission region in the PPV space. Thus it is also investigated.
Table~\ref{tab:flux} summarizes the total numbers of voxels~(Cols. 2-6), total fluxes~(Cols. 7-11) recovered by 
the five data sets, the voxel number ratios and the flux ratios of MWISP-to-OGS~(Cols. 12-13), and 
MWISP-to-CfA~(Cols. 14-15) in the case of full resolution and same resolution, respectively.
Here the total flux is simply the integration of all the emission: F= $\Sigma_i\,T_i\delta\theta_x\delta\theta_y\delta\,v$. 

We find that the voxel number and flux recovered by the MWISP survey are always the largest. In the case of different resolutions,
the voxel number ratio is meaningless and thus is not compared here.
The total flux recovered by the raw MWISP survey is about 1.3 times larger than that of the OGS survey and 1.1 times 
larger than that of the CfA survey. Even considering the uncertainty possibly introduced by calibration, with 
typically of 10\%, the difference between the MWISP and OGS surveys is still significant.
To better understand which factor is dominant in the discrepancy,
we also compared the results identified by surveys with the fixed resolutions but different sensitivities.
We find that the total fluxes recovered by the smoothed MWISP data sets (Cols. 8-9) are somewhat larger than
that from the raw data (Col. 7). 
Naturally, the total flux ratios of MWISP-to-OGS and MWISP-to-CfA
are both shifted to an even higher value of 1.6 in the case of the same resolution.

As previously mentioned, the \co~ observation covers a large range of volumes of the Milky Way~(out to Galactocentric
radii of $R$=$\sim$25~kpc).
The observational effects would be expected
to increase with distance as decreasing the detection limits of the cloud size and mass.
Thus, we also examine the statistical results across the four spiral arms roughly defined 
according to the \vlsr~ distribution (mentioned in \S~\ref{sec:lv}).
Note that the inter-arm regions are ignored here given their negligible contributions.
It is obvious that both the voxel number and flux drop rapidly beyond the Perseus arm, which 
differs by several orders of magnitude in different arms.
This is particularly severe by the CfA and OGS survey. Consequently, 
the voxel number ratios and flux ratios rise to as high as 27.9 and 7.4 for MWISP-to-OGS,
and 53.3 and 43.8 for MWISP-to-CfA in the farthest OSC arm.
In addition, we see that the voxel number ratios are always larger than the flux ratios for all arm components.
The voxel number ratios and flux ratios across the four spiral arms are also shown in Figure~\ref{fig:ratio}.

\begin{deluxetable}{cccccccccccccccccc}
\tabletypesize{\scriptsize}
\setlength{\tabcolsep}{0.05in}
\tablewidth{0pt}
\tablecaption{Total number of voxel and total flux recovered.\label{tab:flux}}
\tablehead{
Arm & \multicolumn{5}{c}{Total voxel} &&  \multicolumn{5}{c}{Total flux} && \multicolumn{5}{c}{Voxel ratio\&Flux ratio}  \\
 \cline{2-6}  \cline{8-12} \cline{14-18} 
 &  \multicolumn{3}{c}{MWISP} & OGS & CfA && \multicolumn{3}{c}{MWISP} & OGS & CfA  &&  \multicolumn{2}{c}{MWISP/OGS} &&\multicolumn{2}{c}{MWISP/CfA} \\
\cline{2-4}  \cline{8-10}  \cline{14-15} \cline{17-18} 
&  raw & sm1 & sm2 & raw & raw &&  raw & sm1 & sm2 & raw & raw  &&  raw & sm1 && raw & sm2  \\
  (1) & (2) & (3) & (4) & (5) & (6) & & (7)&(8)&(9)&(10)&(11)&& (12) & (13) && (14) & (15) }
\startdata
All  & 4.2e7 &  5.4e6  & 1.5e5 & 1.4e6 & 9.0e4 &&   5.2e6 & 6.3e6 & 7.7e6 & 3.9e6 & 4.9e6 && --\&1.3  & 3.9\&1.6 && --\&1.1 & 1.7\&1.6 \\
\hline
Loc  & 3.2e7 &  4.2e6  & 1.1e5 & 1.0e6 & 6.7e4 &&   3.9e6 & 4.7e6 & 5.9e6 & 2.7e6 & 3.7e6 && --\&1.4  & 4.2\&1.7 && --\&1.1 & 1.6\&1.6 \\
Per  & 8.6e6 &  1.1e6  & 3.7e4 & 3.3e5 & 2.1e4 &&   1.3e6 & 1.5e6 & 1.7e6 & 1.1e6 & 1.2e6 && --\&1.2  & 3.3\&1.4 && --\&1.1 & 1.8\&1.4 \\
Out  & 5.0e5 &  8.8e4  & 4.2e3 & 1.0e4 & 1.6e3 &&   4.9e4 & 6.7e4 & 1.0e5 & 2.5e4 & 5.0e4 && --\&2.0  & 8.8\&2.7 && --\&1.0 & 2.6\&2.0 \\
OSC  & 1.8e4 &  3.9e3  & 1.6e2 & 1.4e2 & 3.0   &&   1.7e3 & 2.6e3 & 2.8e3 & 3.5e2 & 6.4e1 && --\&4.9  &27.9\&7.4 && --\&26.6& 53.3\&43.8\\
\enddata
\tablecomments{Columns (2)-(6): total number of voxels recovered by the raw MWISP survey, the MWISP survey with the same
resolutions as the OGS and CfA surveys, the raw OGS and CfA surveys, respectively. 
Columns (7)-(11): total flux of the recovered voxels, in units of \kkmsarc.
Columns (12)-(13): voxel ratio and flux ratio of MWISP/OGS, with the original resolution and same resolution, respectively.
Columns (14)-(15): voxel ratio and flux ratio of MWISP/CfA, with the original resolution and same resolution, respectively.
}
\end{deluxetable}

\subsection{Completeness of the Flux \label{sec:completeness}}
The fluxes recovered, $F\,$($>$$T_{\rm MB}$), by each survey above a given main beam temperature, $T_{\rm MB}$, 
are plotted in Figure~\ref{fig:flux}, both in the case of the original resolution and sampling~(left panel) 
and the same resolution and sampling~(middle and right panels). 
The values of $T_{\rm MB}$ range from $T_{\rm MB}=$3$\sigma_{\rm rms}$
~(near the lowest cloud boundary as defined in \S\ref{sec:identify}, that is, sensitivity limit) to the 
peak temperature of the cloud, $T_{\rm peak}$.
To some extent, the high end $T_{\rm MB}$ is mainly affected by the resolutions and sampling of the observations, 
and we would expect its distribution to move to somewhat higher values of $T_{\rm peak}$ with observations characterized by 
better resolutions and sampling. Indeed, the MWISP and OGS surveys with higher resolutions have much higher 
dynamic ranges of $T_{\rm MB}$ than the CfA survey.
In addition, we find that at a given $T_{\rm MB}$, the fluxes $F\,$($>$$T_{\rm MB}$) recovered by 
the MWISP survey are larger than those by the CfA survey, while are slightly less than those by the OGS survey  
even compared under the same angular and velocity resolutions. 

\begin{deluxetable*}{cccccccccccccccc}
\tabletypesize{\scriptsize}
\setlength{\tabcolsep}{0.07in}
\tablewidth{0pt}
\tablecaption{Flux completeness.\label{tab:completeness}}
\tablehead{
Arm & \multicolumn{5}{c}{Parameter $a$} &&  \multicolumn{5}{c}{Parameter $b$} && \multicolumn{3}{c}{Completeness}  \\
\cline{2-6}  \cline{8-12} \cline{14-16}
 &  \multicolumn{3}{c}{MWISP} & OGS & CfA && \multicolumn{3}{c}{MWISP} & OGS & CfA  &&  \multicolumn{3}{c}{MWISP} \\
\cline{2-4}  \cline{8-10}  \cline{14-16}
 &  raw & sm1  &  sm2  & raw & raw &&  raw & sm1 & sm2  & raw & raw  &&  raw & sm1 & sm2  \\
 (1) & (2) & (3) & (4) & (5) & (6) && (7)&(8)&(9)&(10)&(11)&& (12) & (13) & (14)  }
\startdata
All & 9.0e6 & 8.3e6 & 8.2e6 & 1.0e7 & 5.9e6 &&0.38 & 0.38 & 0.58 & 0.34 & 0.87 && 0.58 & 0.75 & 0.94 \\
\hline
Loc & 7.5e6 & 6.7e6 & 6.4e6 & 9.7e6 & 4.6e6 &&0.44 & 0.44 & 0.62 & 0.42 & 0.95 && 0.52 & 0.71 & 0.91 \\
Per & 1.8e6 & 1.7e6 & 1.7e6 & 2.0e6 & 1.2e6 &&0.25 & 0.25 & 0.52 & 0.22 & 0.63 && 0.70 & 0.85 & 1.01 \\
Out & 1.3e5 & 1.2e5 & 1.2e5 & 1.0e5 & 9.9e4 &&0.70 & 0.72 & 2.17 & 0.50 & 3.10 && 0.38 & 0.61 & 0.80 \\
OSC & 5.3e3 & 5.3e3 & 6.0e3 & 1.2e3 & 1.9e2 &&0.84 & 0.95 & 6.51 & 0.43 & 5.00 && 0.32 & 0.54 & 0.44 \\
\enddata
\tablecomments{We fit the $F\,$($>$$T_{\rm MB}$)-$T_{\rm MB}$ relation with an 
exponential function, $F\,$($>$$T_{\rm MB})=a\,{\rm exp}(-bT_{\rm MB})$, where $a$ and $b$ are two parameters.
The derived values for $a$ and $b$ are summarized in Columns (2)-(6) and Columns (7)-(11), respectively. 
The flux completeness, $F\,$($>$3$\sigma$)/$F\,$($>$0), for the MWISP survey with the original resolution (Col. 12) 
and same resolutions as the OGS (Col. 13) and CfA (Col. 14) surveys are also tabulated.
}
\end{deluxetable*}

Figure~\ref{fig:flux} shows a nonlinear falling of $F\,$($>$$T_{\rm MB}$) with $T_{\rm MB}$ for all data sets, which can be approximately described by 
an exponential function, $F\,$($>$$T_{\rm MB})=a\,{\rm exp}(-b\,T_{\rm MB})$, where $a$ and $b$ are two parameters. 
$a$ is equivalent to the flux that we would expect to measure with perfect sensitivity (a theoretical 
$T_{\rm MB}$=0~K isosurface). A larger value of $b$ implies a faster decreasing of $F\,$($>$$T_{\rm MB}$) with $T_{\rm MB}$.
We use \texttt{curve\_fit} in the \texttt{Python} package \texttt{SciPy} to fit the distribution, with flux errors considered. 
The error of recovered flux is estimated with $\sqrt N \sigma$, where $N$ is the voxel number, and $\sigma$ is the rms noise of each voxel. 
The derived relations are marked in each panel, which are apparently different across a variety of data sets.
The values of $a$ and $b$ both seem to strongly depend on the spectral and spatial resolutions of the data. 
Indeed, when different data sets are smoothed to the same resolution, consistent 
values of both $a$ and $b$ are derived (refer to middle and right panels of Fig.~\ref{fig:flux}). 
To avoid chaos, only the exponential fitting results to the MWISP data are overlapped. 

Using these relations, we can also extrapolate the flux to
the $T_{\rm MB}$=0 K isosurface. Hence, 
the completeness of flux can be calculated by the ratio of the
flux at 3$\sigma$ cutoff, $F\,$($>$3$\sigma$), and the expected flux given infinite sensitivity, $F\,$($>$0~K).
We see that the estimated flux completeness is also suffered observational effects, as higher expected values of $F\,$($>$0~K),
while somewhat lower values of $F\,$($>$3$\sigma$) are found in the data sets with higher resolutions. 
Thus the corresponding completeness estimated here might only represent
an upper limit, particularly in the data set with relatively coarse resolution, at which the majority of clouds are
commonly marginally resolved or detected.

The flux recovered and flux completeness are also investigated in the four spiral arms with 
different Galactocentric radii (refer to Fig.~\ref{fig:lpoo}).
In the top panel, the results are derived from data with the original resolution and sampling.
In comparison, the middle and bottom panels are in the case of the same resolution and sampling. 
We see that the conclusions drawn from the whole region in Fig.~\ref{fig:flux} are still valid in these four sub-regions. 
The MWISP and OGS data have a similar $b$ value for a given arm component, which is smaller than that of the CfA data. 
We find that the value of $b$ parameter seems to increase rapidly beyond the Perseus arm, which implies a rapid 
falling of $F\,$($>$$T_{\rm MB}$) with $T_{\rm MB}$. 
As expected, the flux recovered is less complete in the more distant spiral arm, which is only 32\% in the 
OSC arm, even by the MWISP survey. 
It is important to note that the estimation only represents a lower limit, as previously discussed. 
The derived $F\,$($>$$T_{\rm MB})$-$T_{\rm MB}$ relations and the flux completeness are all summarized in Table~\ref{tab:completeness}.
\subsection{Recovered Molecular Clouds and Their Area and Mass\label{sec:number}}
In the following two subsections, the molecular gas properties are derived based on MC samples, i.e., 
the whole sample and the sub-samples in the four individual spiral arms.
MCs are assigned to each spiral arm based on \vlsr~range as mentioned in \S~\ref{sec:lv}.
We use the parallax-based distance estimator of \citet{reid2019} to determine the distance
to each cloud. Given the large uncertainties of kinematic distances to the MCs in the Local arm, 
these MCs are excluded in the following analysis.
The typical distances for the whole sample and the three sub-samples are summarized 
in Table~\ref{tab:distance}. The molecular gas masses are derived
by assuming a constant CO-to-H$_{\rm 2}$ conversion factor of 2.0$\times$10$^{\rm 20}$~cm$^{\rm -2}{\rm (K\,km\,s^{-1})^{-1}}$~\citep{bolatto2013}.
Following \citet{yan2020}, the MC's properties are measured. Table~\ref{tab:number} summarizes the 
total number of MCs~(Cols. 2-6), the total projected angular area of MCs~(Cols. 7-11),
the total mass of MCs~(Cols. 12-16), and the area ratio and mass ratio~(Cols. 17-20) that recovered
by the five data sets of the three surveys.

\begin{deluxetable*}{cccccccccccccccccccc}
\tabletypesize{\scriptsize}
\setlength{\tabcolsep}{0.04in}
\tablewidth{0pt}
\tablecaption{Summary of the molecular cloud distances.\label{tab:distance}}
\tablehead{
& \multicolumn{5}{c}{Median} && \multicolumn{5}{c}{Mean} &&\multicolumn{5}{c}{S.D.}   \\
\cline{2-6}  \cline{8-12} \cline{14-18}
& \multicolumn{3}{c}{MWISP} & OGS & CfA &&\multicolumn{3}{c}{MWISP} & OGS & CfA &&  \multicolumn{3}{c}{MWISP} & OGS & CfA \\
\cline{2-4}  \cline{8-10}  \cline{14-16}
&  raw & sm1  & sm2  & raw & raw &&  raw & sm1  & sm2  & raw & raw && raw & sm1  & sm2  & raw & raw  \\
(1) & (2) & (3) & (4)& (5) & (6) && (7) &(8)&(9)&(10) & (11) &&(12)&(13)& (14)&(15)&(16)  }
\startdata
All &  2.67 &  2.69 &  2.69 &  2.67 &  2.66 &&  3.35 &  3.53 &  3.90 &  3.05 &  3.33 &&  1.79 &  1.93 &  2.41 &  1.35 &  1.92 \\
\hline
Per &  2.64 &  2.64 &  2.54 &  2.66 &  2.62 &&  2.77 &  2.80 &  2.71 &  2.76 &  2.67 &&  0.87 &  0.92 &  0.87 &  0.81 &  0.72 \\
Out &  5.96 &  6.06 &  6.24 &  6.05 &  5.96 &&  5.96 &  6.24 &  6.54 &  5.78 &  6.20 &&  1.64 &  1.22 &  1.32 &  1.59 &  1.42 \\
OSC & 10.36 & 10.29 & 10.29 & 10.57 & 17.37 && 10.63 & 10.60 & 10.58 & 11.38 & 17.37 &&  1.74 &  1.82 &  2.58 &  2.34 &  --  \\
\enddata
\tablecomments{
The statistical parameters for MC’s distance (in units of kpc) are summarized, including the median value~(Columns 2-6), 
mean value (Columns 7-11), and standard deviation (Columns 12-16) for the MWISP raw data, same resolutions as OGS 
and CfA surveys, raw OGS and CfA surveys, respectively.
}
\end{deluxetable*}

\begin{deluxetable*}{cccccccccccccccccccccccc}
\tabletypesize{\scriptsize}
\setlength{\tabcolsep}{0.04in}
\tablewidth{0pt}
\tablecaption{Summary of the molecular cloud properties.\label{tab:number}}
\tablehead{
Arm & \multicolumn{5}{c}{Total number} &&  \multicolumn{5}{c}{Total area} && \multicolumn{5}{c}{Total mass} && \multicolumn{5}{c}{Area ratio\&Mass ratio}  \\
 \cline{2-6}  \cline{8-12} \cline{14-18} \cline{20-24}
    &  \multicolumn{3}{c}{MWISP} & OGS & CfA && \multicolumn{3}{c}{MWISP} & OGS & CfA &&\multicolumn{3}{c}{MWISP} & OGS & CfA &&  \multicolumn{2}{c}{MWISP/OGS} &&\multicolumn{2}{c}{MWISP/CfA} \\
\cline{2-4}  \cline{8-10}  \cline{14-16}  \cline{20-21} \cline{23-24}
 &  raw & sm1 & sm2 & raw & raw &&  raw & sm1 & sm2 & raw & raw && raw & sm1 & sm2 & raw & raw &&  raw & sm1 && raw & sm2  \\
 (1) & (2) & (3) & (4) & (5) & (6) & & (7)&(8)&(9)&(10)&(11)& &(12)&(13)&(14)&(15)&(16)&& (17) & (18) && (19) & (20) }
\startdata
All &  7368 &  4599 &   221 &  1946 &   156 && 65.0  & 75.0  & 1.9e2 & 27.0  & 1.2e2 && 5.2e6 & 6.1e6 & 5.6e6 & 4.2e6 & 4.0e6 && 2.4\&1.3 & 2.7\&1.5 && 0.5\&1.3 & 1.6\&1.4  \\ 
\hline
Per &  6231 &  3752 &   163 &  1773 &   130 && 57.8  & 65.9  & 1.7e2 & 26.0  & 1.1e2 && 4.3e6 & 4.8e6 & 4.6e6 &3.7e6 & 3.2e6 && 2.2\&1.1 & 2.5\&1.3 && 0.5\&1.3 & 1.6\&1.4  \\
Out &  1007 &   747 &    48 &   165 &    25 && 6.6   & 8.2   & 19.7  & 1.4   & 1.2e1 && 8.4e5 & 1.2e6 & 8.1e5 &4.2e5 & 7.4e5 && 4.8\&2.0 & 6.0\&2.9 && 0.5\&1.1 & 1.6\&1.1  \\
OSC &   130 &   100 &    10 &     8 &     1 && 0.4   & 0.6   & 2.5   & 0.02  & 0.08  && 1.2e5 & 1.6e5 & 1.6e5 &2.6e4 & 2.1e4 && 19.4\&4.6 & 26.1\&6.1 && 5.5\&5.6 & 31.8\&7.6 \\
\enddata
\tablecomments{Columns (2)-(6): total number of objects recovered by the MWISP data with original resolution, same
resolutions as OGS, and CfA surveys, raw OGS and CfA surveys, respectively.
Columns (7)-(11): the total projected area of the recovered objects, in units of deg$^{\rm 2}$.
Columns (12)-(16): the total mass of the recovered objects, in units of $M_{\sun}$.
Columns (17)-(18): area ratio and mass ratio of MWISP/OGS, with the original resolution and same resolution, respectively.
Columns (19)-(20): area ratio and mass ratio of MWISP/CfA, with the original resolution and same resolution, respectively.
}
\end{deluxetable*}

We find that the number of molecular clouds identified by the MWISP survey with the full resolution is up to $\sim$2 orders of magnitude
larger than that of the CfA survey, and $\sim$3 times larger than that of the OGS survey.
Such significant discrepancy is perhaps not surprising because the recovered objects are
controlled by the sensitivity and resolution of the data set.
As stated in \S~\ref{sec:identify}, regions of emission could be successfully recovered as single entities only if their sizes are larger than
the resolutions of the observations, the separations of the blend sources are larger than the resolutions,
and their brightness temperatures are above a user-defined threshold.
Additionally, the sensitivity of the data set that defines the boundary of the molecular cloud also has somewhat effects
on the decomposition results, because the higher sensitivity data would tend to merge multiple MCs into a single MC.
When MWISP data are smoothed and resampled to the resolutions and pixel sizes of the OGS and CfA surveys,
we find that comparable numbers of clouds to the OGS and CfA surveys are identified in these smoothed MWISP data, with values of 8133 and 286.

The total projected area of MCs recovered by the raw MWISP is 2.0 times larger than that recovered by the OGS survey, 
but is only 80 per cent of that recovered by the CfA survey. 
Here the deconvolution is performed by subtracting the beam from the measured projected area.
It seems that the smoothing of the MWISP data to the OGS velocity resolution does not significantly change the projected area recovered
(75 deg$^{\rm 2}$ that is similar to the raw MWISP). Thus the area ratio of the smoothed 
MWISP to OGS almost remains unchanged as compared to that of the raw data.
However, the smoothing of the MWISP to the CfA resolution produces a significant boost of projected area recovered
(up to 190 deg$^{\rm 2}$ that is 2.9 times larger than that of the raw MWISP), as a consequence, 
the area ratio of the smoothed MWISP to CfA increases up to 1.6.

As expected, the total molecular gas mass of the MCs recovered by the MWISP is the highest, which is 1.5 and 1.4 
times those recovered by the OGS and CfA surveys in the case of the full resolution, respectively.   
We also examine the statistical results recovered across the three sub-samples.
It shows that the number, projected area, and mass of the clouds recovered all decrease rapidly
beyond the Perseus arm, particularly significant in results recovered by the CfA and OGS surveys.
As a consequence, the area ratios and mass ratios of MWISP-to-CfA and MWISP-to-OGS rise with increasing distance~(plotted in Fig.~\ref{fig:ratio}).

\subsection{Distributions of physical parameters}

\begin{deluxetable*}{cccccccccccccccccccccccccc}
\tabletypesize{\scriptsize}
\setlength{\tabcolsep}{0.04in}
\tablewidth{0pt}
\tablecaption{Statistical parameters of the molecular cloud physical size.\label{tab:size}}
\tablehead{
& \multicolumn{5}{c}{Min} && \multicolumn{5}{c}{Max} &&\multicolumn{5}{c}{Median} &&\multicolumn{5}{c}{Mean}  \\
\cline{2-6}  \cline{8-12} \cline{14-18} \cline{20-24}
&  \multicolumn{3}{c}{MWISP} & OGS & CfA && \multicolumn{3}{c}{MWISP} & OGS & CfA &&\multicolumn{3}{c}{MWISP} & OGS & CfA &&  \multicolumn{3}{c}{MWISP} & OGS & CfA \\
\cline{2-4}  \cline{8-10}  \cline{14-16}  \cline{20-22} 
  &  raw & sm1 & sm2 & raw & raw &&  raw & sm1 & sm2 & raw & raw && raw & sm1 & sm2 & raw & raw &&  raw & sm1 & sm2 & raw & raw  \\
(1) & (2) & (3) & (4)& (5) & (6) && (7) &(8)&(9)&(10) & (11) &&(12)&(13)& (14)&(15)&(16)&&(17)&(18)&(19)&(20)&(21)  }
\startdata
All &  0.3 &  0.6 &  6.4 &  0.7 &  5.4 && 64.8 & 75.2 & 170.1 & 40.3 & 138.5 &&  1.3 &  1.9 & 16.4 &  1.8 & 14.7 &&  2.0 &  2.9 & 23.6 &  2.5 & 19.9 \\
\hline
Per &  0.3 &  0.6 &  6.4 &  0.7 &  5.4 && 64.8 & 75.2 & 170.1 & 40.3 & 138.5 &&  1.1 &  1.6 & 12.3 &  1.7 & 11.8 &&  1.7 &  2.4 & 18.0 &  2.4 & 16.5 \\
Out &  0.5 &  1.0 & 10.4 &  1.1 & 14.6 && 31.4 & 57.6 & 101.6 & 14.6 & 78.0  &&  2.7 &  3.7 & 33.9 &  3.7 & 33.6 &&  3.6 &  5.0 & 36.6 &  4.5 & 36.3  \\
OSC &  1.9 &  2.1 & 26.2 &  4.0 & 47.8 && 24.4 & 27.3 & 76.3  & 10.9 & 47.8  &&  4.6 &  6.4 & 51.5 &  5.4 & 47.8 &&  5.3 &  7.2 & 51.1 &  5.9 & 47.8  \\
\enddata
\tablecomments{The statistical parameters for MC’s size (in units of pc) are summarized, including the minimum~(Columns 2-6), maximum (Columns 7-11), 
median~(Columns 12-16), and mean (Columns 17-21) values, for the MWISP raw data, same resolutions as OGS and CfA surveys, raw OGS and CfA surveys, respectively.
}
\end{deluxetable*}

\begin{deluxetable*}{cccccccccccccccccccccccccc}
\tabletypesize{\scriptsize}
\setlength{\tabcolsep}{0.04in}
\tablewidth{0pt}
\tablecaption{Statistical parameters of the molecular cloud mass.\label{tab:mass}}
\tablehead{
& \multicolumn{5}{c}{Min} && \multicolumn{5}{c}{Max} &&\multicolumn{5}{c}{Median} &&\multicolumn{5}{c}{Mean}  \\
\cline{2-6}  \cline{8-12} \cline{14-18} \cline{20-24}
&  \multicolumn{3}{c}{MWISP} & OGS & CfA && \multicolumn{3}{c}{MWISP} & OGS & CfA &&\multicolumn{3}{c}{MWISP} & OGS & CfA &&  \multicolumn{3}{c}{MWISP} & OGS & CfA \\
\cline{2-4}  \cline{8-10}  \cline{14-16}  \cline{20-22}
&  raw & sm1 & sm2 & raw & raw &&  raw & sm1 & sm2 & raw & raw && raw & sm1 & sm2 & raw & raw &&  raw & sm1 & sm2 & raw & raw  \\
(1) & (2) & (3) & (4)& (5) & (6) && (7) &(8)&(9)&(10) & (11) &&(12)&(13)& (14)&(15)&(16)&&(17)&(18)&(19)&(20)&(21)  }
\startdata
All & 1.0  & 6.4  & 1.4e2 & 25.0  & 2.7e2 && 1.1e6 & 1.3e6 & 2.6e6 & 5.9e5 & 1.9e6 && 34.0  & 95.0  & 2.8e3 & 2.7e2 & 3.2e3 && 7.1e2 & 1.3e3 & 2.5e4 & 2.1e3 & 2.6e4 \\
\hline
Per & 1.0  & 6.4  & 1.4e2 & 25.0  & 2.7e2 && 1.1e6 & 1.3e6 & 2.6e6 & 5.9e5 & 1.9e6 && 26.0  & 66.0  & 1.5e3 & 2.4e2 & 2.2e3 && 6.8e2 & 1.3e3 & 2.8e4 & 2.1e3 & 2.5e4 \\
Out & 2.7  & 19.0 & 9.3e2 & 55.0  & 2.0e3 && 6.6e4 & 1.9e5 & 1.7e5 & 4.1e4 & 1.6e5 && 1.2e2 & 2.9e2 & 7.4e3 & 9.4e2 & 1.8e4 && 8.3e2 & 1.6e3 & 1.7e4 & 2.5e3 & 3.0e4 \\
OSC & 34.0 &1.0e2 & 5.6e3 & 9.1e2 & 2.1e4 && 2.3e4 & 2.9e4 & 4.8e4 & 1.0e4 & 2.1e4 && 3.1e2 & 7.9e2 & 1.3e4 & 1.8e3 & 2.1e4 && 9.1e2 & 1.6e3 & 1.6e4 & 3.2e3 & 2.1e4 \\
\enddata
\tablecomments{The statistical parameters for MC’s mass (in units of $M_{\sun}$) are summarized, including the minimum~(Columns 2-6), maximum (Columns 7-11), 
median~(Columns 12-16), and mean (Columns 17-21) values, for the MWISP raw data, same resolutions as OGS and CfA surveys, raw OGS and CfA surveys, respectively.
}
\end{deluxetable*}

\begin{deluxetable*}{cccccccccccccccccccccccccc}
\tabletypesize{\scriptsize}
\setlength{\tabcolsep}{0.04in}
\tablewidth{0pt}
\tablecaption{Statistical parameters of the molecular cloud surface density.\label{tab:surface}}
\tablehead{
& \multicolumn{5}{c}{Min} && \multicolumn{5}{c}{Max} &&\multicolumn{5}{c}{Median} &&\multicolumn{5}{c}{Mean}  \\
\cline{2-6}  \cline{8-12} \cline{14-18} \cline{20-24}
&  \multicolumn{3}{c}{MWISP} & OGS & CfA && \multicolumn{3}{c}{MWISP} & OGS & CfA &&\multicolumn{3}{c}{MWISP} & OGS & CfA &&  \multicolumn{3}{c}{MWISP} & OGS & CfA \\
\cline{2-4}  \cline{8-10}  \cline{14-16}  \cline{20-22}
&  raw & sm1 & sm2 & raw & raw &&  raw & sm1 & sm2 & raw & raw && raw & sm1 & sm2 & raw & raw &&  raw & sm1 & sm2 & raw & raw  \\
(1) & (2) & (3) & (4)& (5) & (6) && (7) &(8)&(9)&(10) & (11) &&(12)&(13)& (14)&(15)&(16)&&(17)&(18)&(19)&(20)&(21)  }
\startdata
All &  1.5 &  2.5 &  0.8 & 10.7 &  2.0 && 82.4 & 73.9 & 28.4 & 256.9 & 32.0 &&  6.2 &  7.6 &  2.8 & 26.2 &  4.7 &&  7.9 &  9.3 &  3.7 & 29.9 &  5.7  \\
\hline
Per &  1.5 &  2.5 &  0.8 & 10.7 &  2.0 && 82.4 & 73.9 & 28.4 & 256.9 & 32.0 &&  6.4 &  7.9 &  3.1 & 26.5 &  5.0 &&  8.2 &  9.7 &  4.0 & 30.4 &  6.0  \\
Out &  1.8 &  3.0 &  0.8 & 13.2 &  2.4 && 27.9 & 27.4 & 10.5 & 66.1 &  9.0  &&  5.6 &  6.7 &  2.4 & 23.7 &  4.2 &&  6.5 &  7.5 &  3.0 & 24.9 &  4.7  \\
OSC &  2.3 &  3.1 &  0.8 & 16.3 &  2.9 && 21.9 & 20.1 &  2.6 & 34.8 &  2.9  &&  5.1 &  6.0 &  1.6 & 26.2 &  2.9 &&  6.1 &  7.0 &  1.8 & 25.0 &  2.9  \\
\enddata
\tablecomments{The statistical parameters for MC’s surface density (in units of $M_{\sun}$$\,$pc$^{\rm -2}$) are summarized, including the minimum~(Columns 2-6), 
maximum (Columns 7-11), median~(Columns 12-16), and mean (Columns 17-21) values, for the MWISP raw data, same resolutions as OGS and CfA surveys, 
raw OGS and CfA surveys, respectively.}
\end{deluxetable*}

The statistical parameters for MC's size, mass, mass surface density are summarized in Tables~\ref{tab:size}-\ref{tab:surface},
including the minimum, maximum, median, and mean values for each property.
We find that all parameters presented here are different between the five data sets, which seem to be 
controlled by the data quality and the distance to MC. 
As expected, the MWISP MC sample shows the lowest statistical values of mass surface density, as MWISP data are capable of detecting 
faint emission around the periphery of the cloud and a large number of small and faint clouds that were largely missed by 
previous surveys.  

Figure~\ref{fig:prop} presents the cumulative distributions of the velocity dispersion ($\sigma_{\rm v}$), effective radius ($R_{\rm eff}$), 
and mass ($M$) for the molecular clouds extracted from the five data sets. Both the whole cloud sample,
and the sub-samples in the Perseus, Outer, and OSC arms are compared. 
For a better comparison, each property is plotted in figures with a fixed range of scale.
It shows that both the dynamic range of each property and the shape of each distribution are significantly different between the 
five data sets of the three surveys across the Galactic disk. The dynamic ranges of all these three fundamental properties exhibit the largest range
in the MWISP data, while the smallest range in the CfA data. 
It seems that those small or fain MCs could only be detected by those surveys with high resolution and high sensitivity.
We quantify the difference 
by applying both a truncated power-law function, $N(x>x{'})=N_{0}[(x/x_{0})^{\gamma+1}-1]$, and a non-truncated power-law function, 
$N(x>x{'})=[(x/x_{0})^{\gamma+1}]$~\citep[e.g.,][]{heyer2015,rice2016,colombo2019} to the cloud size and mass distributions. 
Here the completeness limits are evaluated by the mean size and mean mass of objects recovered by each data set~(refer to Tables~\ref{tab:size}-\ref{tab:mass}), 
which represent conservative estimations. 

Using the maximum-likelihood method of \citet{rosolowsky2005} and \citet{rosolowsky2007}, 
the size and mass spectra are measured.
We find that the size and mass distributions in the outer Galaxy are better fitted 
by non-truncated power-law functions in most situations.
The only exception is the size spectrum of the whole cloud sample extracted from the raw MWISP data, which
can not be well fitted at the high-value end. 
The MCs' size function fitting results for the different data sets and different distance components 
are summarized in Table~\ref{tab:fit_size}. Similarly, the MCs' mass function fitting results are summarized in Table~\ref{tab:fit_mass}.
Note that the fitting results for clouds number below 10 have large uncertainties thus are not fitted here.
%

As an example, Figure~\ref{fig:fit} shows the fitting results of the size and mass distributions of 
the whole cloud sample extracted from the MWISP data with full resolution.   
The MCs' size distribution derives an index of $\gamma$=$-$2.93$\pm$0.05, and the mass distribution derives an index of 
$\gamma$=$-$1.83$\pm$0.05. The index of the mass spectrum is very similar to the results
for catalogs in the first Galactic quadrant extracted by using the same algorithms and input parameters, as well as the 
same CO survey data, e.g., with $\gamma$=$-$1.8 reported by \citet{yan2020}~(their Figure 17) for the local MC sample, 
and with $\gamma$=$-$1.74 reported by \citet{su2021} for the MC sample located near the tangent points.

We see that the indices fitted by other data sets with different resolutions and/or sensitivities are quite different from
those of the raw MWISP data. Besides, the exact value of $\gamma$ varies in different distance components. 
Generally, the slight steeper $\gamma$ values for both size and mass distributions are revealed by data with higher resolution 
and low sensitivity, and also by cloud samples with larger distances.
 Nevertheless, all survey data across all distance components display ``bottom heavy" size distributions
 with indices $\gamma<-$2, while ``top heavy" mass distributions with indices $\gamma>-$2.
This implies that although the vast majority of clouds are small ones, 
the majority of the mass in the Galactic plane is confined to the giant ones.
These properties are similar to what have been found in the inner Galaxy~\citep[refer to the review by][]{heyer2015}.

Compared to other studies toward the same Outer Galaxy, we find that our indices of mass are much 
shallower. For examples, the indices reported by \citet{rosolowsky2005} and \citet{rice2016} are both with 
values of $\gamma<-$2. The cloud catalog used by \citet{rosolowsky2005} was directly from 
\citet{heyer2001} on the basis of the OGS survey, while the catalog used by \citet{rice2016} was extracted by 
dendrogram on the basis of the CfA data. Similar to this study, both studies identified objects as contiguous regions
above a fixed intensity threshold, but used data sets with different resolutions, and chose 
different cutoffs and methods for decomposing structures. 

\begin{deluxetable*}{cccccccccccccccccc}
\tabletypesize{\scriptsize}
\setlength{\tabcolsep}{0.04in}
\tablewidth{0pt}
\tablecaption{Parameters of MC size spectra derived from different data sets.\label{tab:fit_size}}
\tablehead{ & \multicolumn{5}{c}{Number} && \multicolumn{5}{c}{$\gamma$} &&\multicolumn{5}{c}{$R_{0}$} \\
\cline{2-6}  \cline{8-12} \cline{14-18}
&  \multicolumn{3}{c}{MWISP} & OGS & CfA && \multicolumn{3}{c}{MWISP} & OGS & CfA &&\multicolumn{3}{c}{MWISP} & OGS & CfA \\
\cline{2-4}  \cline{8-10}  \cline{14-16}  
    &raw & sm1 & sm2 & raw & raw &&  raw & sm1 & sm2 & raw & raw && raw & sm1 & sm2 & raw & raw  \\  
(1) & (2) & (3) & (4)& (5) & (6) && (7) &(8)&(9)&(10) & (11) &&(12)&(13)& (14)&(15)&(16)  }
\startdata
All & 2157 & 1392 & 73 & 589 & 44 &&$-$2.93$\pm$0.05 &$-$2.97$\pm$0.05 &$-$3.01$\pm$0.20 &$-$3.04$\pm$0.06 &$-$2.75$\pm$0.20 && 115$\pm$12 & 120$\pm$12 & 216$\pm$50 & 60$\pm$ 5 & 184$\pm$39 \\
\hline
Per &1780 &1055 &44 &515  &43 &&$-$2.97$\pm$0.05 &$-$2.98$\pm$0.08 &$-$2.57$\pm$0.25 &$-$3.05$\pm$0.07 &$-$3.07$\pm$0.54 &&76$\pm$6 &80$\pm$11 &179$\pm$63 &50$\pm$5  &95$\pm$31 \\
Out &327  &214  &18 &57   &10 &&$-$3.11$\pm$0.11 &$-$3.05$\pm$0.11 &$-$3.72$\pm$0.47 &$-$3.60$\pm$0.32 &$-$3.52$\pm$0.73 &&57$\pm$9  &69$\pm$9 &103$\pm$19 &23$\pm$5  &92$\pm$22 \\
OSC &48   &39   &-- &--   &-- &&$-$3.99$\pm$0.54 &$-$4.18$\pm$0.51 &--             &--             &--             &&20$\pm$4  &23$\pm$4 &--         &--       &--         \\
\enddata
\tablecomments{The number of objects above the completeness limit~(Columns 2-6), power-law index of the size spectrum~(Columns 7-11), 
and the upper size of the distribution~(in units of pc, Columns 12-16) recovered by the five data sets of the three surveys, respectively.}
\end{deluxetable*}
\begin{deluxetable*}{cccccccccccccccccc}
\tabletypesize{\tiny}
\setlength{\tabcolsep}{0.03in}
\tablewidth{0pt}
\tablecaption{Parameters of MC mass spectra derived from different data sets.\label{tab:fit_mass}}
\tablehead{ & \multicolumn{5}{c}{Number} && \multicolumn{5}{c}{$\gamma$} &&\multicolumn{5}{c}{$M_{0}$} \\
\cline{2-6}  \cline{8-12} \cline{14-18}
&  \multicolumn{3}{c}{MWISP} & OGS & CfA && \multicolumn{3}{c}{MWISP} & OGS & CfA &&\multicolumn{3}{c}{MWISP} & OGS & CfA \\
\cline{2-4}  \cline{8-10}  \cline{14-16}
&raw & sm1 & sm2 & raw & raw &&  raw & sm1 & sm2 & raw & raw && raw & sm1 & sm2 & raw & raw  \\  
(1) & (2) & (3) & (4)& (5) & (6) && (7) &(8)&(9)&(10) & (11) &&(12)&(13)& (14)&(15)&(16)  }
\startdata
All & 505 & 353 & 24 & 203 & 14 &&$-$1.83$\pm$0.05 &$-$1.84$\pm$0.05 &$-$1.88$\pm$0.22 &$-$1.99$\pm$0.13 &$-$1.64$\pm$0.21 &&(1.3$\pm$0.5)e6 & (1.4$\pm$0.6)e6 & (8.7$\pm$6.3)e5 & (4.7$\pm$2.6)e5 & (1.5$\pm$0.8)e6 \\
\hline
Per &345 & 222 &10  &158 &-- && $-$1.83$\pm$0.24& $-$1.83$\pm$0.29&$-$1.71$\pm$0.20  &$-$1.93$\pm$0.07 &--            &&(7.5$\pm$6.4)e5&(8.5$\pm$4.7)e5&(1.2$\pm$1.1)e6&(4.8$\pm$2.6)e5&--             \\
Out &123 & 95  &14  &38  &-- && $-$1.87$\pm$0.16& $-$1.95$\pm$0.20& $-$2.91$\pm$0.71 &$-$2.21$\pm$0.18 &--            &&(2.1$\pm$2.2)e5&(2.0$\pm$1.5)e5&(8.1$\pm$5.0)e4&(5.8$\pm$2.5)e5&-- \\    
OSC &29  & 25  &--  &--  &-- && $-$2.25$\pm$0.46& $-$2.60$\pm$0.64& --             & --           &--            &&(1.4$\pm$0.5)e4&(1.2$\pm$0.8)e4& --           &--             &-- \\
\enddata
\tablecomments{The number of objects above the completeness limit~(Columns 2-6), power-law index of the mass spectrum~(Columns 7-11), 
and the upper mass of the distribution~(in units of $M_{\sun}$, Columns 12-16) recovered by the five data sets of the three surveys, respectively.}
\end{deluxetable*}
		  %
%
%

\section{Discussions}
\subsection{Lessons from the Current CO Surveys}


On the basis of three independent large-scale CO survey data with a variety of resolutions, sensitivities, and observation strategies, 
we compare the global or integrated properties of the molecular clouds, particularly the total flux recovered from these surveys.
Since our analyses are employed by using a uniform decomposition method toward the exact same region, 
it is straightforward to estimate the observational effects on the molecular cloud properties.
To isolate the effects of resolutions on the measured properties, the higher resolution MWISP data are smoothed to
the same spatial and/or velocity resolutions as the OGS and CfA data.
To better understand the variety of observational effects at different distances, we also 
compare the measured properties derived across different spiral arms along the line of sight.  

In practical observations, $T_{\rm MB}$ is clipped due to the finite sensitivity. The recovered flux 
is therefore controlled by the sensitivity of observation~\citep[][]{yan2021b}. 
Furthermore, many studies have suggested that a significant fraction of the observed molecular gas 
is emitted from regions with low temperature and low column density~\citep[e.g.,][]{heyer1998,goldsmith2008,pineda2008,sun2020}. 
All these naturally predict a significant boost of recovered flux with improving 
sensitivity. As expected, the smoothing of MWISP data to the velocity and/or spatial resolutions
of the OGS and CfA surveys decreases the noise level~(Tab.~\ref{tab:survey}), which in turn results 
in increasing of recovered flux~(refer to Cols. 7-9 of Tab.~\ref{tab:flux}).
For the same reason, the significant sensitivity difference between the MWISP and OGS surveys~(a factor of 4.1) 
in the case of same resolution likely leads to the high MWISP collected flux, which exceeds that of OGS by a factor of 1.6. 
Similarly, the recovered flux ratio of MWISP-to-CfA resembles that of the sensitivity ratio~(about 1.7).
Therefore, our statistical comparisons may lend support to the view that the recovered flux is mainly governed 
by the sensitivity of the observation, as stated at the beginning. 

On the other hand, due to the finite resolution (or beam size) and inhomogeneous property of CO emission, 
the observed brightness temperature, $T_{\rm MB}$ will be inevitably diminished by instruments~\citep[][]{yan2021b}. 
We see that larger dynamic ranges of $T_{\rm MB}$ are generally revealed by surveys having 
higher resolutions (left panel of Fig.~\ref{fig:flux}). In addition, we find that the dynamic ranges of 
$T_{\rm MB}$ for all surveys are decreasing from the nearby spiral arm to the distant spiral arm~(Fig.~\ref{fig:lpoo}).
This can be well explained by the decreasing angular size of the emission region, and therefore 
the decreasing of beam filling factor with increasing of distance for each survey.
All these findings are in agreement with the previous study about resolution effects, as mentioned at the beginning. 

For each individual survey, we find that the level of sensitivity clip effect changes in different spiral arms.
This might be attributed to the changing of S/N ratios in different spiral arms, as the decreasing of $T_{\rm MB}$ 
with increasing distance but not the noise level. All of these effects discussed here have to be kept in 
mind when comparing cloud samples suffered from different levels of observational effects.

Additionally, we find that the produced cloud catalogs are dependent on the resolution and sensitivity of the observation.
The data sets with coarse resolution often produce catalogs of clouds with much smaller numbers but larger projected areas
by blending adjacent clouds and missing the clouds with size below the resolution element of the observation.
Thus the size and mass distributions of clouds are shallower than the true distributions.
The observational effects on the produced cloud catalogs will be researched in near future by Q.-Z. Yan et al. (2021, in preparation).


\subsection{Correction Factor of H$_{\rm 2}$ Mass of the Galaxy}
The total H$_{\rm 2}$ mass of our Galaxy was estimated to be (1.0$\pm$0.3)$\times$10$^{\rm 9}$ $M_{\sun}$, which 
includes a total H$_{\rm 2}$ mass beyond the solar circle of 2.7$\times$10$^{\rm 8}$ $M_{\sun}$. 
These results were based on analyses of the CfA survey data, and assumed 
$X_{\rm CO}$=2.0$\times$10$^{\rm 20}$~cm$^{\rm -2}{\rm (K\,km\,s^{-1})^{-1}}$ and $R_{\sun}$=8.5~kpc~\citep[refer to the review by][]{heyer2015}.  
Unlike the CfA survey that fairly covers the entire Galactic plane, the MWISP survey only covers the northern Galactic plane. 
Therefore, it is not straightforward to estimate the total H$_{\rm 2}$ mass of our Galaxy only based on the MWISP data.

In this study, we compare the total flux recovered from the MWISP and CfA surveys in the second Galactic quadrant, 
which reveals a flux ratio of 1.6. Considering the fact that it is the inner Galaxy that makes the 
largest contributions to the H$_{\rm 2}$ mass of the Galaxy, we also examine the recovered flux ratio of the MWISP 
and CfA surveys in the overlapping region of the first Galactic quadrant~($l$ = [25\fdg8, 49\fdg7], and $b$=[$-$5\fdg03, +5\fdg03]).
The comparison reveals that the flux ratio of the MWISP and CfA surveys is 1.4 in the first Galactic quadrant.
For the third Galactic quadrant, the flux ratio of these two surveys is not being investigated at the moment, 
yet is expected to be much higher than those found in the first and second Galactic quadrants due to much more severe observational effects.
Flux is, to the zero-order of magnitude, proportional to the H$_{\rm 2}$ mass.
Thus the flux ratio could approximately indicate the total recovered H$_{\rm 2}$ mass ratio by the two surveys. 
Actually, the total mass ratio of the tabulated clouds of the MWISP and CfA surveys is derived to be 1.4 in \S~\ref{sec:number}, 
which is indeed close to the flux ratio. The discrepancy between the flux and mass ratios might be mainly contributed by the 
slight discrepancy of distance derived by different cloud catalogs, which can be found in Table~\ref{tab:distance}.

The application of these ratios of MWISP and CfA to the total H$_{\rm 2}$ mass of our Galaxy leads to a correction factor 
of at least 1.4, which means that the Galaxy should be at least 40\% more massive than previously determined,
if the same $X_{\rm CO}$ and rotation curve were assumed. 
The comparison of mass spectra from various surveys reveals that the index (refer to Tab. 10) 
derived from the MWISP survey ($\gamma$=$-$1.83) is steeper than the CfA survey ($\gamma$=$-$1.64), 
also suggesting a higher total H$_{\rm 2}$ mass based on the MWISP survey.

Furthermore, the flux completeness was estimated to be in the range of 32\%-70\% across the disk even based on the 
improved CO data from the MWISP survey~(refer to \S~\ref{sec:completeness}).
It is important to note that the molecular gas at the distance of $>$20~kpc is outside the current scope of the MWISP survey.
Because in these regions, a typical cloud can only be marginally detected as a compact source, implying rather low flux completeness.
Therefore, we warn that our knowledge of our Galaxy is still far from complete.
If the completeness correction was considered, an even larger fraction of the matter should be contributed by baryonic matter.
A new instrument with the combination of higher sensitivity and resolution is demanded and will provide a more accurate 
estimation of the H$_{\rm 2}$ mass of the Galaxy and will also benefit a wide range of science.
Such data will also allow us to reveal the integrated properties of clouds at uniform
detection limits of the cloud physical size and mass across the Galactic plane, in other words, with
the same degree of observational effects.

\section{Summary}
We report the MC integrated properties that the DBSCAN algorithms recovered by an ongoing CO survey of the MWISP
survey within Galactic coordinates $l$ = [104\fdg75, 141\fdg54] 
and $b$ = [$-$3\fdg028, 5\fdg007]. To better understand the observational effects, 
we also compare our results to those extracted by a uniform decomposition method from the CfA and OGS surveys. 
While a survey with higher resolution and sensitivity can reveal more detailed internal structures in MCs,
we focus this work on the integrated properties of MCs for comparison.
The properties recovered from a variety of spiral arms having different Galactocentric radii are also compared in detail.
We draw several conclusions from the comparison:

(1) The cloud integrated properties are subject to different observational effects.   
The level of observational effects changes in both different surveys and different spiral arms.
The effects are particularly severe in surveys with low-sensitivity, low-resolution, relatively 
coarse sampling, and in distant regions, since in these cases, the clouds are either marginally 
resolved or have low signal-to-noise ratios. 

(2) The total number of voxel and flux integrated over all spiral arms recovered by the MWISP are 3.9 
and 1.6 times larger than those recovered by the OGS survey, and are 1.7 and 1.6 times larger than 
those recovered by the CfA survey, in the case of the same resolution. These discrepancies are 
increasing with distance. In the outermost OSC arm, the mass ratio and flux ratio for MWISP-to-OGS increase 
up to 27.9 and 7.4, respectively, while increase up to 53.3 and 43.8 for MWISP-to-CfA.

(3) The census of molecular gas in our Galaxy is incomplete, even in the two near spiral arms recovered by the MWISP data, 
which have flux completeness lower than $\sim$50\%-70\%. 

(4) The total H$_{\rm 2}$ mass of the tabulated clouds based on the MWISP survey is 1.5 and 1.4 times larger than 
those based on the OGS and CfA surveys, respectively. These facts may imply that the total H$_{\rm 2}$ mass of our Galaxy should be 
at least 40\% more massive than that previously estimated based on the CfA survey.
 
(5) The mass spectrum in the Galactic outskirts is better described by a non-truncating power-law with $\gamma$=$-$1.83$\pm$0.05,
and an upper mass of $M_0$=(1.3$\pm$0.5)$\times$10$^{\rm 6}$~$M_{\sun}$. 
The index of mass is much shallower than other studies toward the same Outer Galaxy, which reported typical values of $\gamma<-$2.
However, this index is consistent with those found in the inner Galaxy, which are also on the basis of the MWISP data
by using the same criteria and methods for decomposing structures.

\begin{acknowledgments}
We would like to thank other members of the MWISP group, Zhiwei Chen for the support of the data reduction system, Jixian Sun, and Dengrong Lu 
for their support of the MWISP database, Yingjie Li, and Yuehui Ma for their contribution to the observation. We are extremely grateful to all the staff 
members at the PMO 13.7 m telescope, particularly the observer assistants for their long-term support. 
Research presented here relied heavily on the use of the CfA 1.2 m CO survey and the FCRAO 14 m outer Galaxy survey data.
We would like to thank Professor Mark Heyer for providing  helpful comments and suggestions that improved the paper.
MWISP was sponsored by the National Key R\&D Program of China with grant no. 2017YFA0402700 and CAS Key Research Program of Frontier Sciences with grant 
no. QYZDJ-SSW-SLH047. YS is supported by the National Natural Science Foundation of China through grant 11773077 and the Youth Innovation Promotion Association, CAS (2018355). 
JY is supported by the National Natural Science Foundation of China through grant 12041305.
\end{acknowledgments}

\facilities{PMO 13.7 m telescope}
\software{astropy \citep{2013A&A...558A..33A}, SciPy~\citep{virtanen2020}.}

\begin{figure*}
\centering
\includegraphics[angle=0,scale=0.9,bb=10 30 595 641]{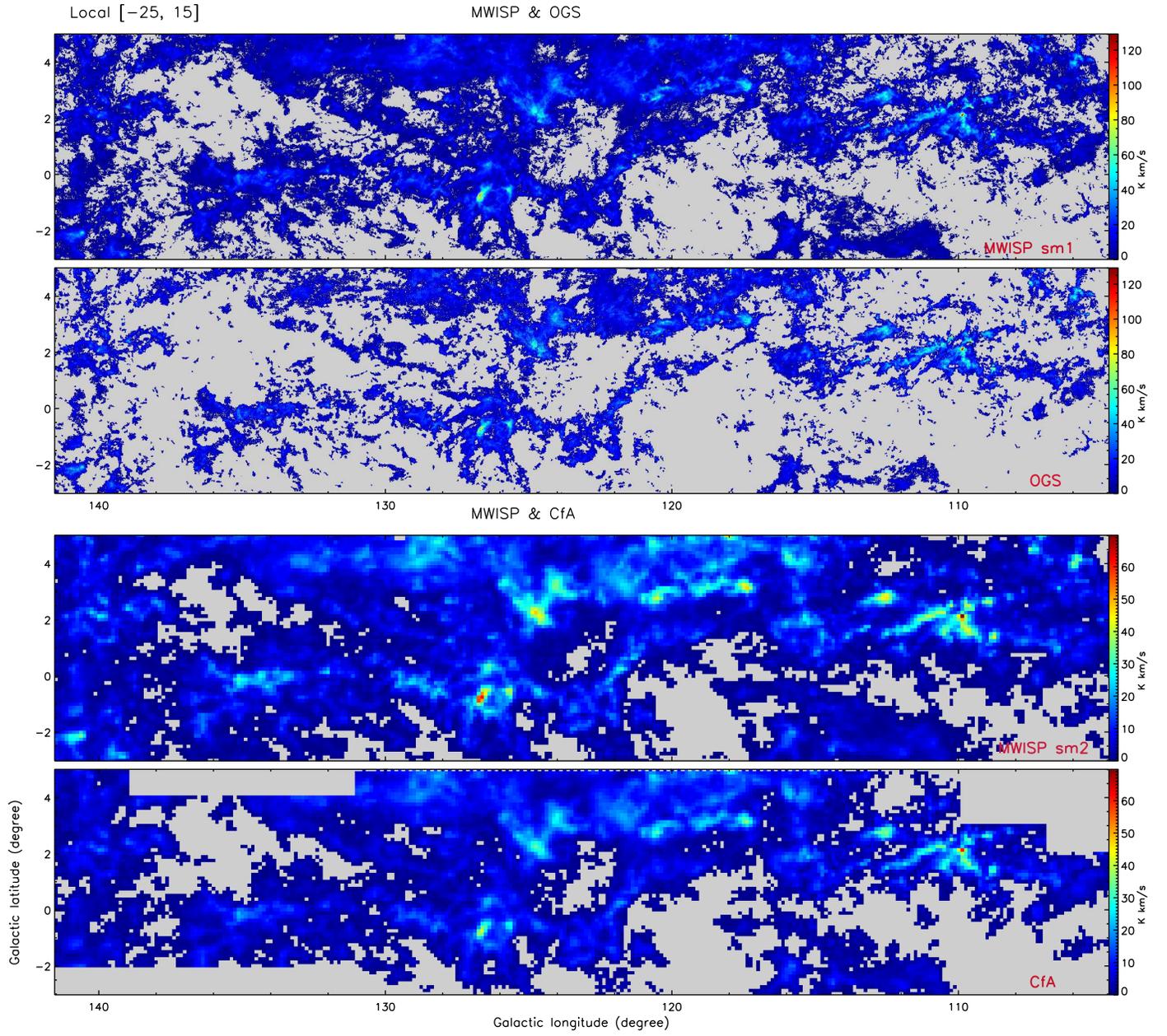}
\vspace{-9mm}
\caption{Integrated intensity maps of \co~ emission. The velocity range is marked in the corner of each map.
The MWISP data are smoothed to match the resolution and sampling of the OGS and CfA surveys.
The images only contain emission identified by DBSCAN. All emission-free pixels are blanked. 
\label{fig:intensity}}
\end{figure*}
\addtocounter{figure}{-1}
\begin{figure*}
\centering
\includegraphics[angle=0,scale=0.9,bb=10 30 595 641]{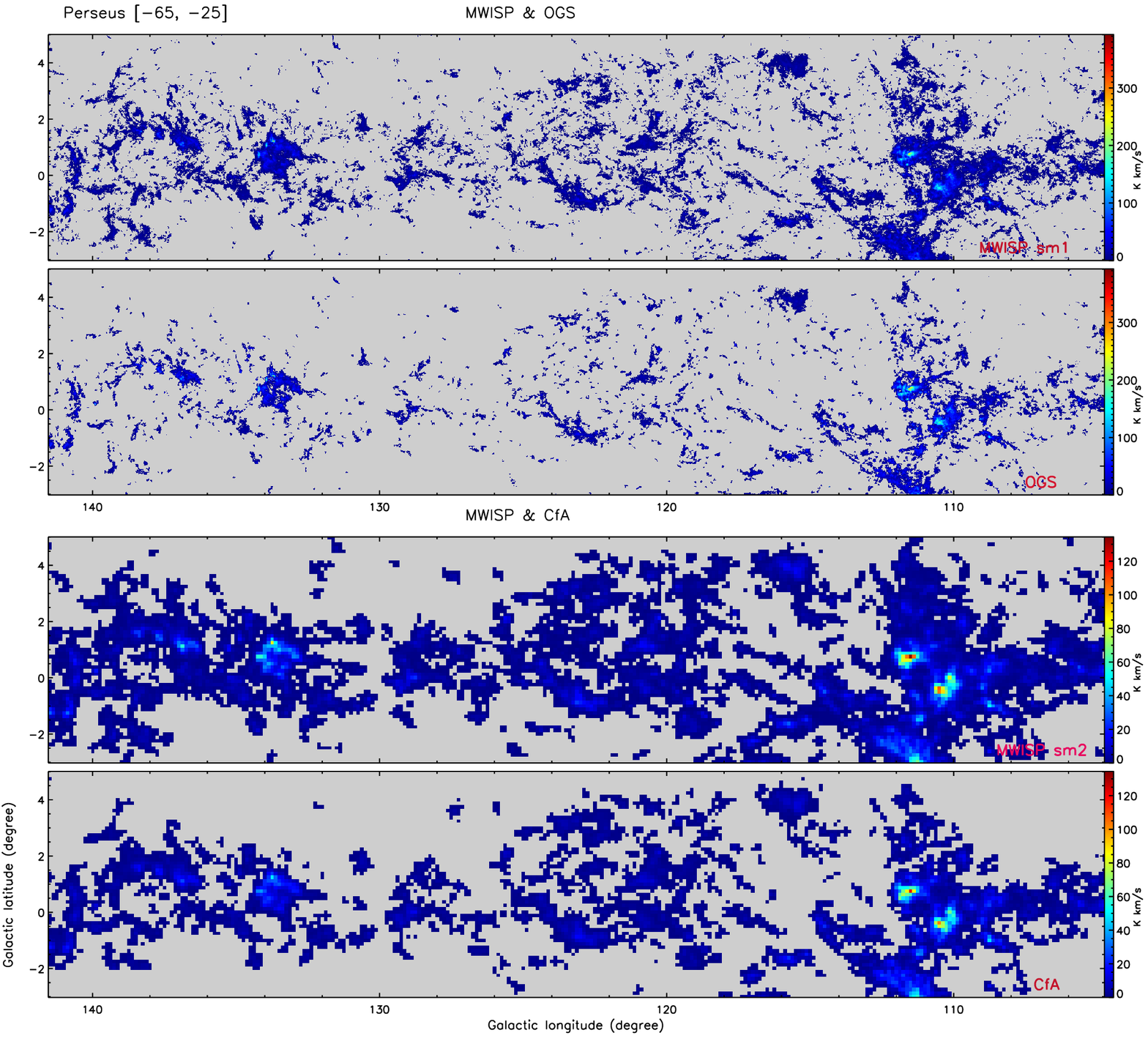}
\vspace{-9mm}
\caption{(Continued).}
\end{figure*}
\addtocounter{figure}{-1}
\begin{figure*}
\centering
\includegraphics[angle=0,scale=0.9,bb=10 30 595 641]{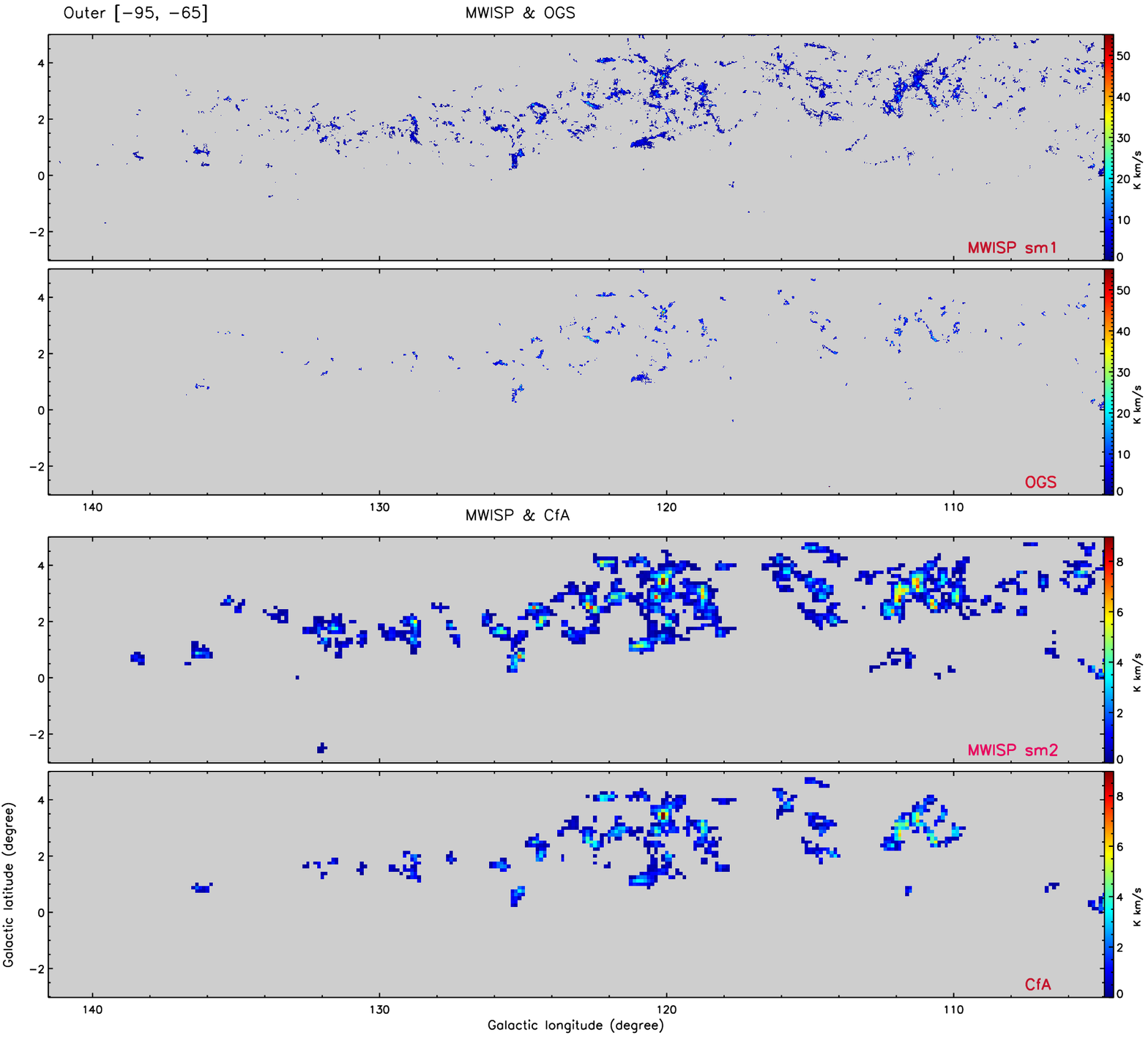}
\vspace{-9mm}
\caption{(Continued).}
\end{figure*}
\addtocounter{figure}{-1}
\begin{figure*}
\centering
\includegraphics[angle=0,scale=0.9,bb=10 30 595 641]{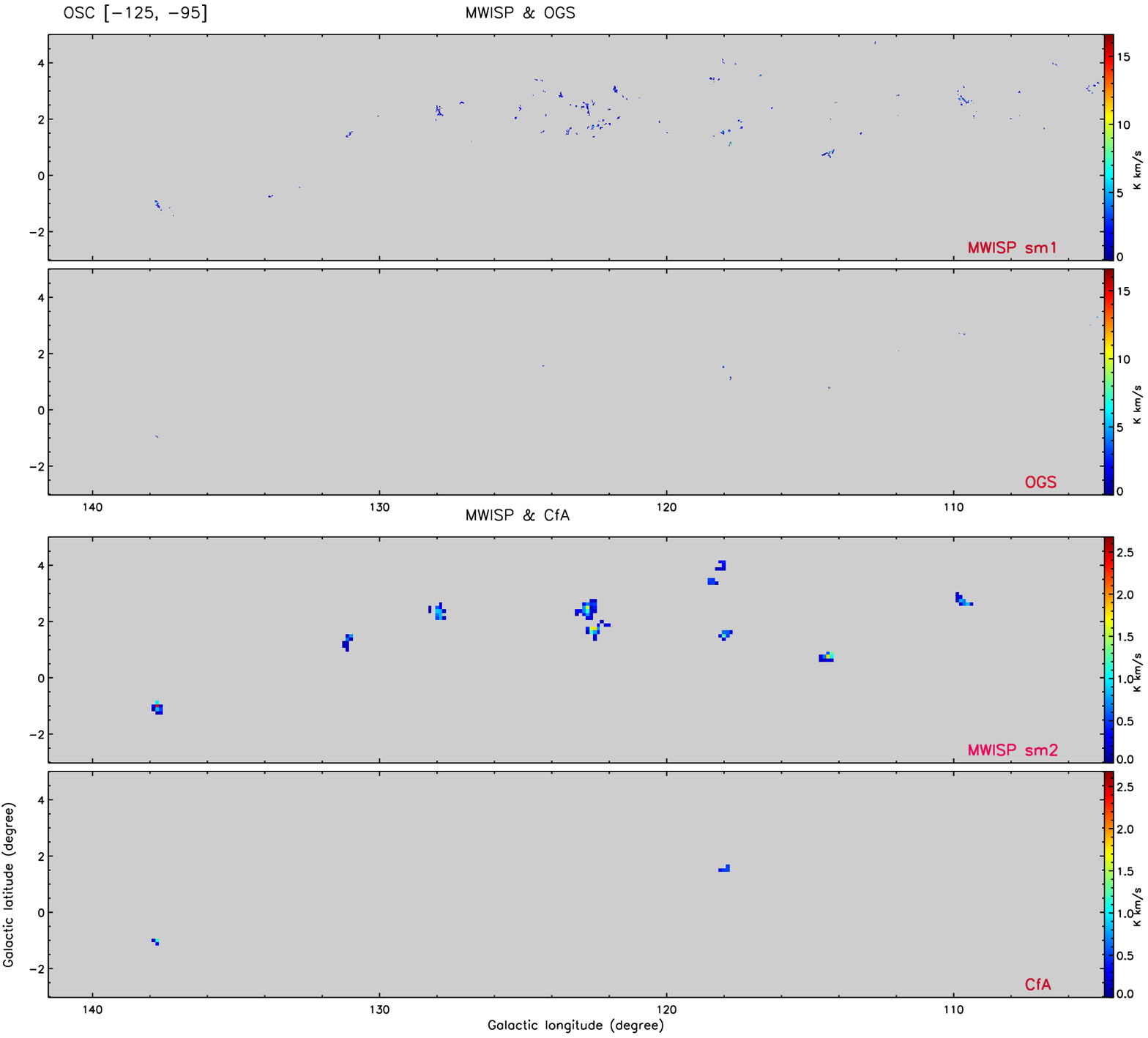}
\vspace{-9mm}
\caption{(Continued).}
\end{figure*}
\begin{figure*}
\vspace{-1mm}
\centering
\includegraphics[angle=0,scale=0.9,bb=70 10 390 300]{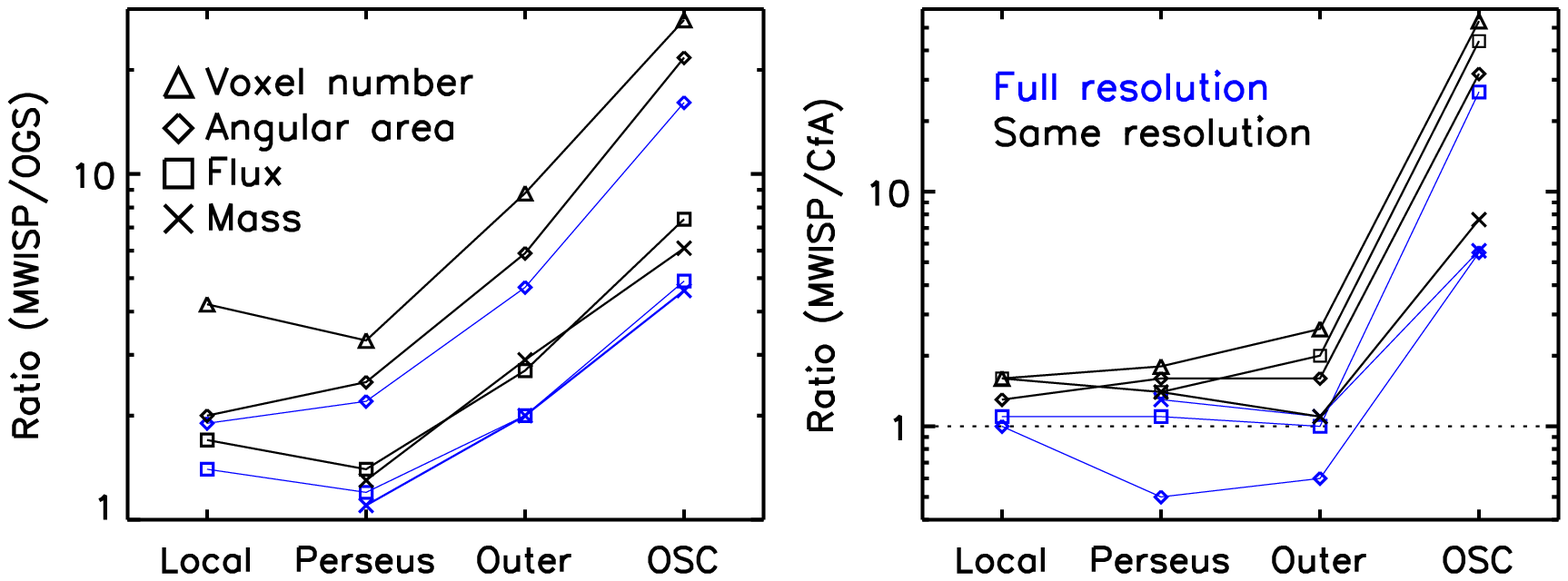}
\caption{($a$) The voxel number ratio, flux ratio, projected area ratio, mass ratio of MWISP-to-OGS in the 
case of full resolution (plotted in blue) and same resolution (plotted in black). ($b$) Same as panel ($a$), 
but for the ratio of MWISP-to-CfA.
\label{fig:ratio}}
\end{figure*}
\begin{figure*}
\centering
\includegraphics[angle=0,bb=-180 260 800 650,scale=0.52]{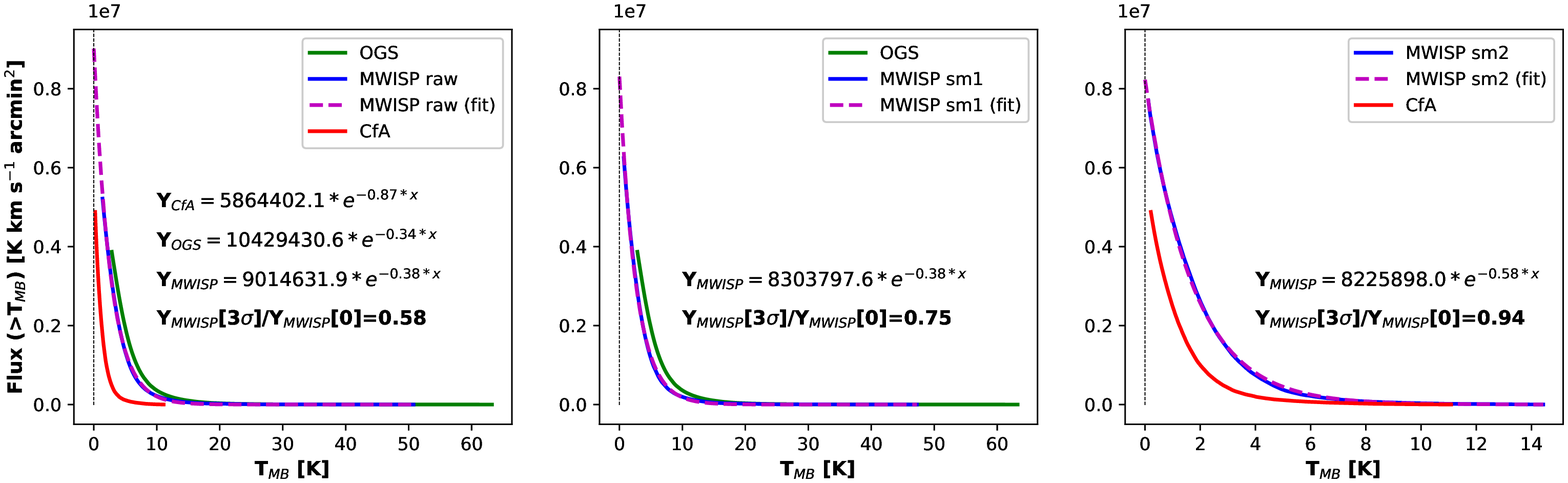}
\caption{The total flux recovered as a function of main beam temperature, $T_{\rm cutoff}$ for the 
MWISP (blue), OGS (green), and CfA (red) surveys. The MWISP data are also smoothed 
to match the resolution and sampling of the OGS data~(middle 
panel), and the CfA data (right panel), respectively. 
\label{fig:flux}}
\end{figure*}
\clearpage
\begin{figure*}
\vspace{13mm}
\centering
\includegraphics[angle=0,scale=0.43,bb=-280 130 1000 450]{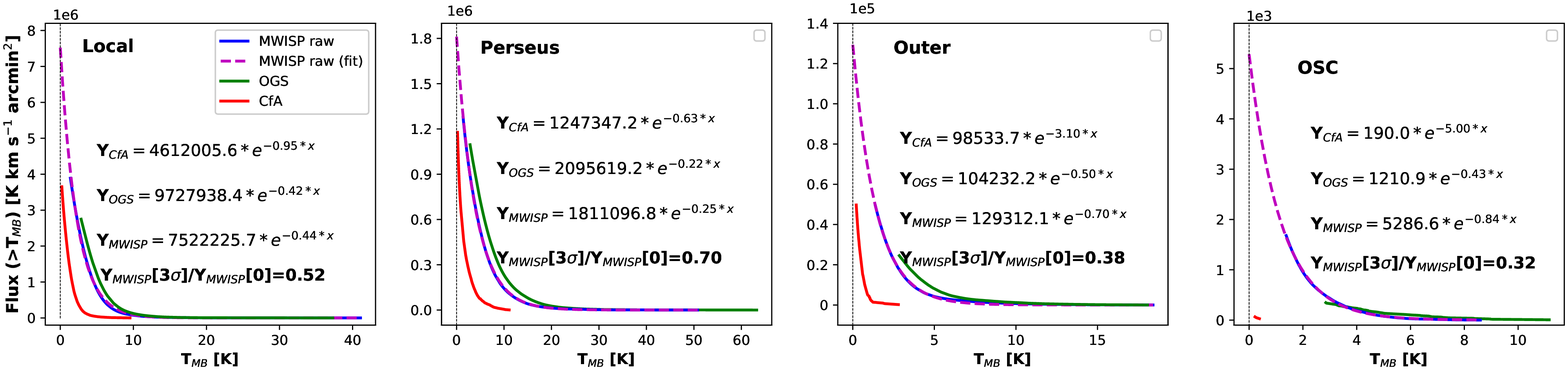}\\
\includegraphics[angle=0,scale=0.43,bb=-280 130 1000 400]{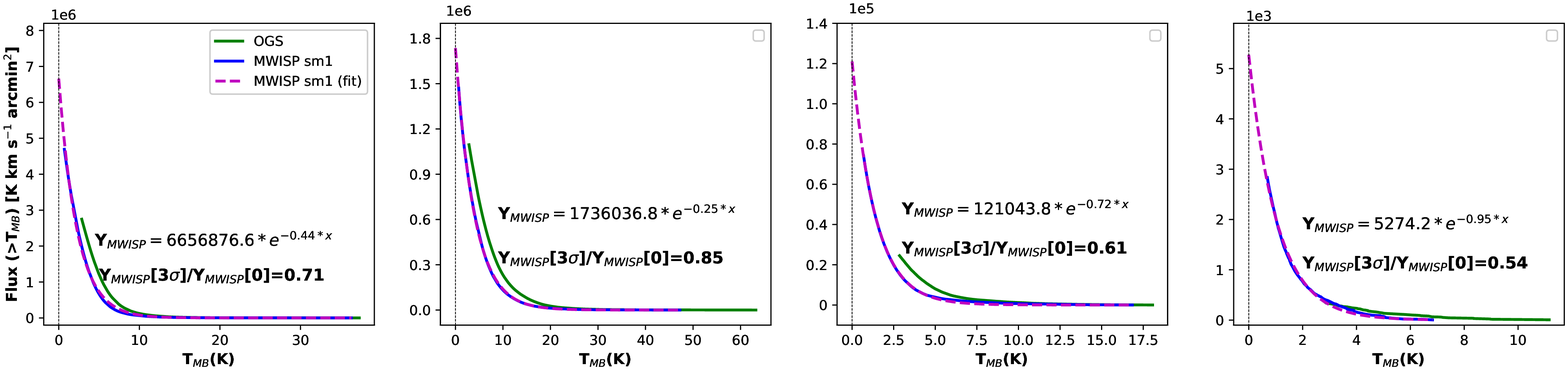}\\
\includegraphics[angle=0,scale=0.43,bb=-280 130 1000 400]{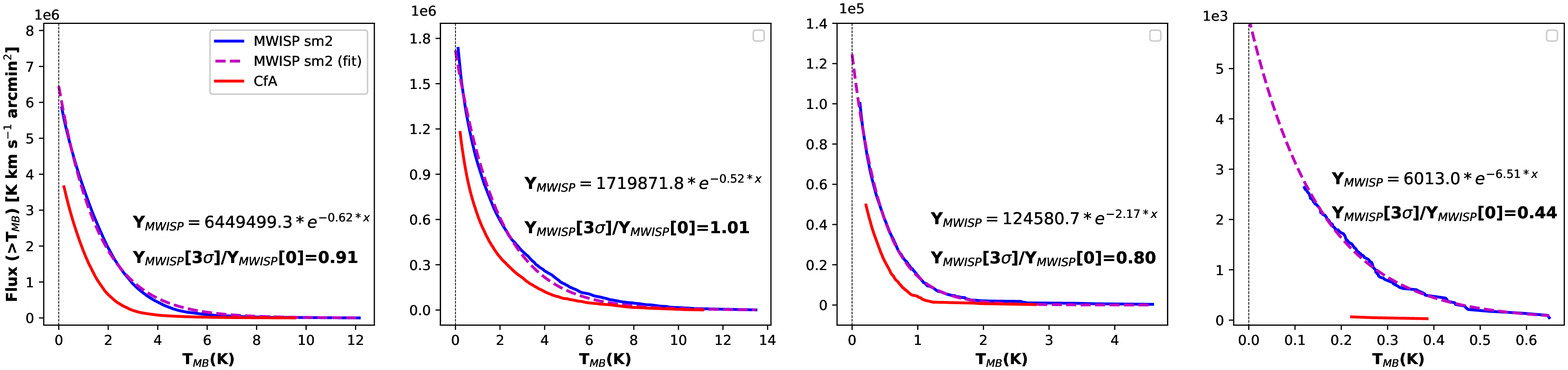}
\caption{Same as Fig.~\ref{fig:flux} but for different spiral arms that are roughly defined according to the
\vlsr~ distributions: (1) the Local arm [$-$25, 15~\kms], (2) the Perseus arm [$-$65, $-$25~\kms],
(3) the Outer arm [$-$95, $-$65~\kms], and (4) the OSC arm [$-$125, $-$95~\kms]. \label{fig:lpoo}}
\end{figure*}

\begin{figure*}
\centering
\includegraphics[angle=0,scale=0.42]{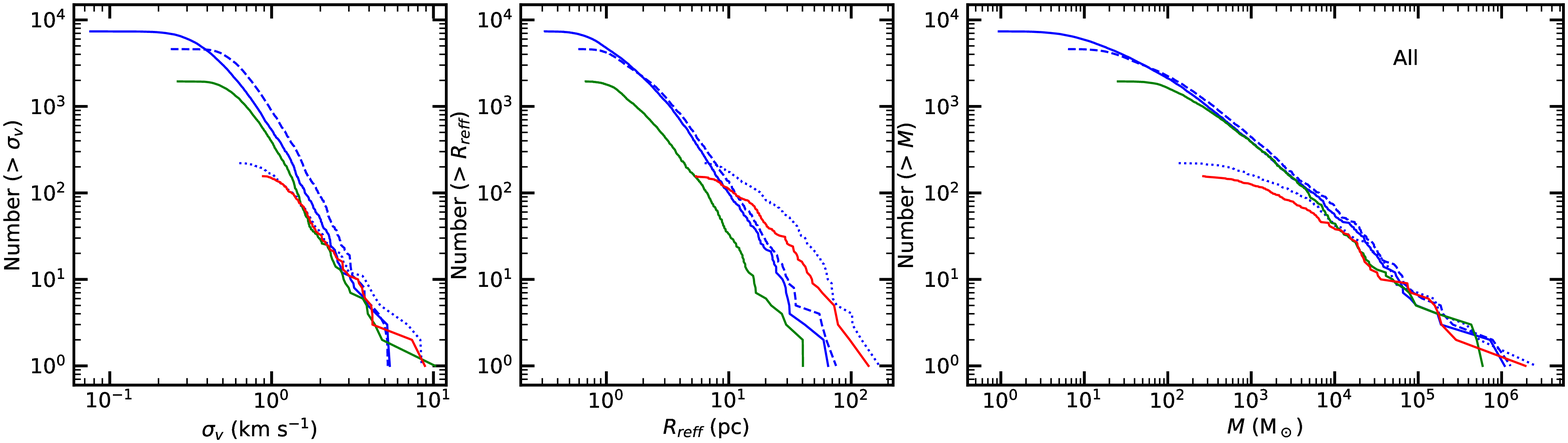}
\includegraphics[angle=0,scale=0.42]{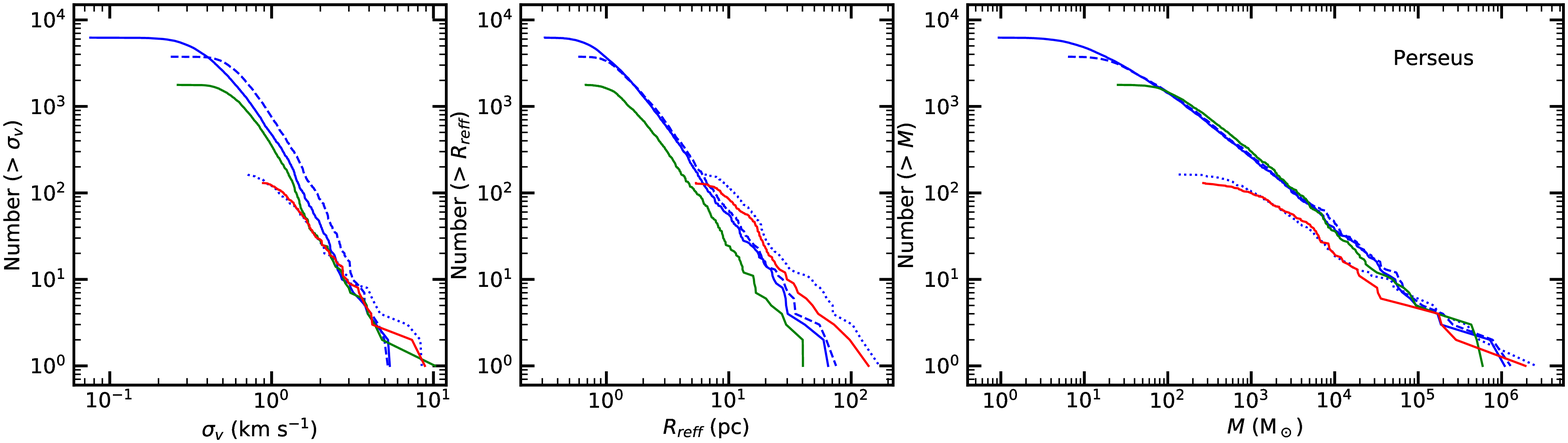}
\includegraphics[angle=0,scale=0.42]{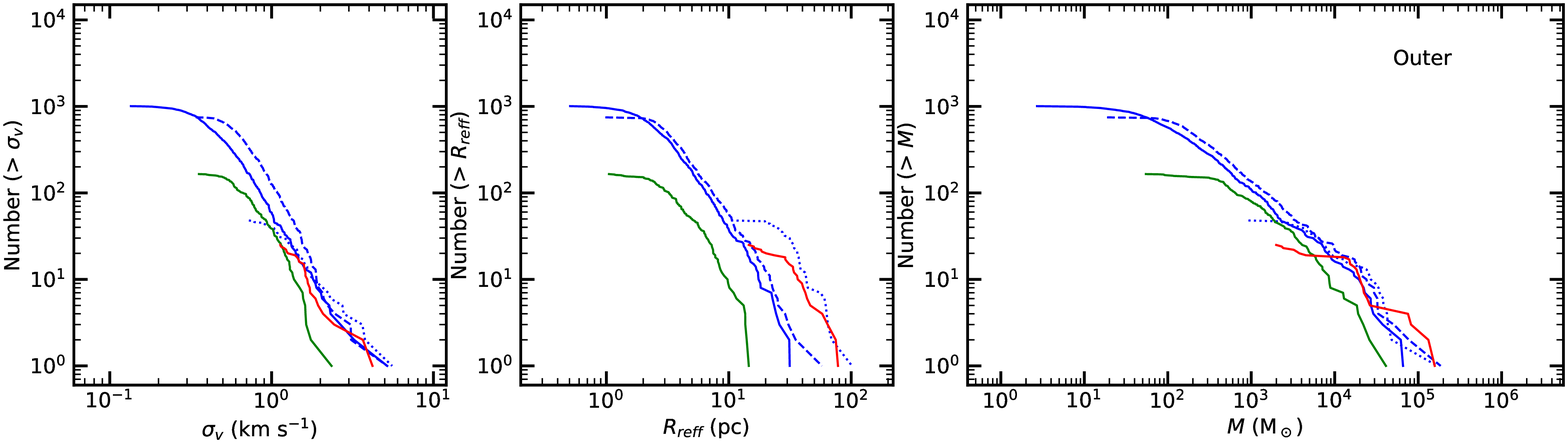}
\includegraphics[angle=0,scale=0.42]{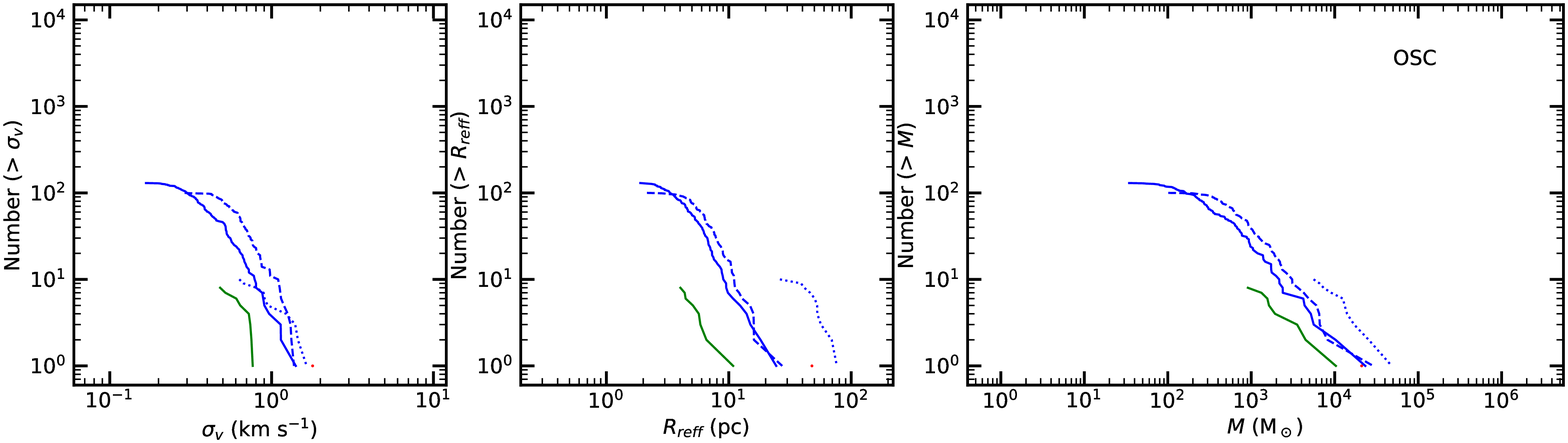}
\caption{The cumulative distributions of velocity dispersion, effective radius, and mass of the molecular clouds that extracted from the 
MWISP (blue), OGS (green), and CfA (red) surveys. The blue-, green-, and red-solid lines indicate the results from the MWISP, OGS, and 
CfA raw data, respectively. The blue-dashed and blue-dotted lines indicate results from the smoothed MWISP data with the same resolution 
as the OGS and CfA data, respectively.~\label{fig:prop}}
\end{figure*}
\begin{figure*}
\centering
\includegraphics[angle=0,scale=0.55]{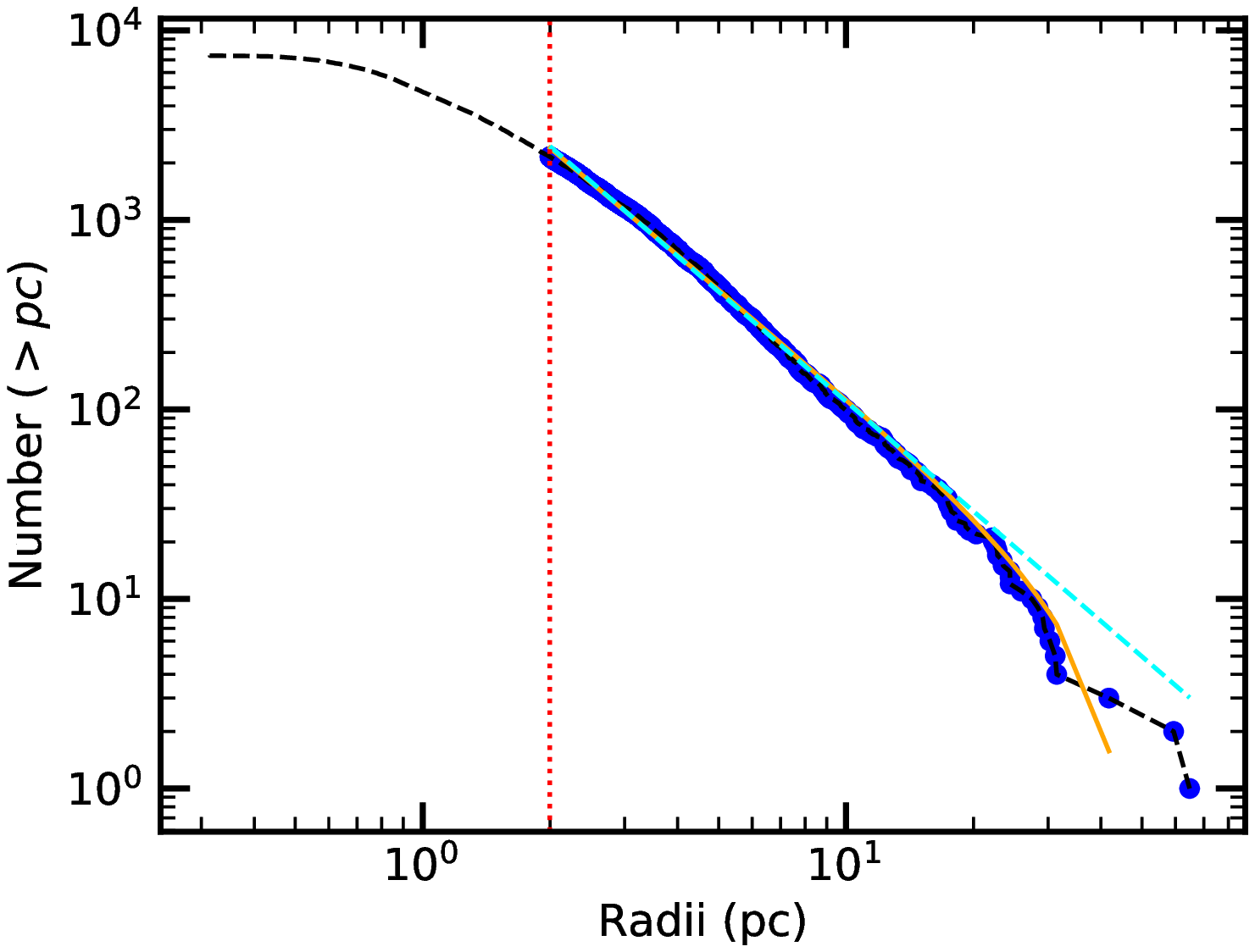}%
\includegraphics[angle=0,scale=0.55]{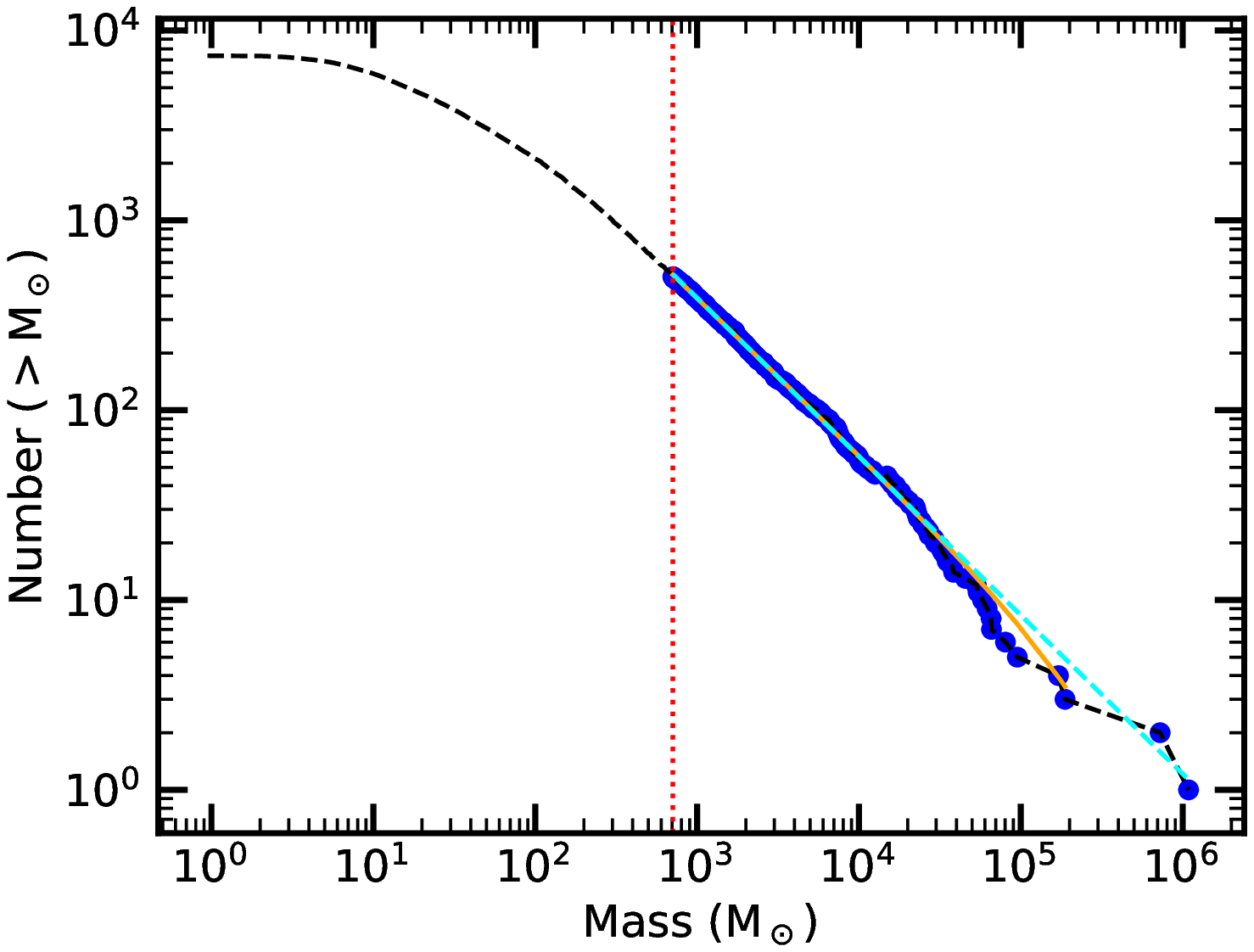}
\caption{Cumulative effective radius (left) and mass (right) functions for the whole cloud sample extracted from the raw MWISP data. 
In each panel, both a truncated (solid orange) and a non-truncated (dashed cyan) power-law spectrum are fit. The 
red-dotted line marks the minimum meaningful value of each property.~\label{fig:fit}}
\end{figure*}
\end{document}